\documentclass[sigconf]{acmart}
\usepackage[ruled, vlined, linesnumbered]{algorithm2e}
\usepackage{setspace}
\usepackage{subcaption}
\usepackage{enumitem} 
\usepackage{multirow}
\usepackage{color}
\usepackage{amsmath}
\usepackage[linesnumbered,ruled]{algorithm2e}
\usepackage{cleveref}
\usepackage{amsmath}
\usepackage{algorithmic}
\usepackage[normalem]{ulem}

\begin{document}

\title{Learning Structure and Knowledge Aware Representation with Large Language Models for Concept Recommendation}

\author{Qingyao Li}
\email{ly890306@sjtu.edu.cn}
\affiliation{
 \institution{Shanghai Jiao Tong University}
  \city{Shanghai}
  \country{China}
}

\author{Wei Xia}
\email{xiawei24@huawei.com}
\affiliation{%
  \institution{Huawei Noah's Ark Lab}
  \city{Shenzhen}
  \country{China}
}
\author{Kounianhua Du}
\email{kounianhuadu@sjtu.edu.cn}
\affiliation{
 \institution{Shanghai Jiao Tong University}
  \city{Shanghai}
  \country{China}
}
\author{Qiji Zhang}
\email{zqjpeter@sjtu.edu.cn}
\affiliation{
 \institution{Shanghai Jiao Tong University}
  \city{Shanghai}
  \country{China}
}

\author{Weinan Zhang}
\email{wnzhang@sjtu.edu.cn}
\affiliation{
 \institution{Shanghai Jiao Tong University}
  \city{Shanghai}
  \country{China}
}

\author{Ruiming Tang}
\email{tangruiming@huawei.com}
\affiliation{
 \institution{Huawei Noah's Ark Lab}
  \city{Shenzhen}
  \country{China}
}
\author{Yong Yu}
\authornote{Corresponding author.}
\email{yyu@apex.sjtu.edu.cn}
\affiliation{
 \institution{Shanghai Jiao Tong University}
  \city{Shanghai}
  \country{China}
}


\newcommand{\our}{\mbox{SKarRec}\xspace}
\renewcommand{\shortauthors}{Qingyao Li, et al.}
\settopmatter{printacmref=false} 
\renewcommand\footnotetextcopyrightpermission[1]{} 

\begin{abstract}


Concept recommendation aims to suggest the next concept for learners to study based on their knowledge states and the human knowledge system. While knowledge states can be predicted using knowledge tracing models, previous approaches have not effectively integrated the human knowledge system into the process of designing these educational models.
In the era of rapidly evolving Large Language Models (LLMs), many fields have begun using LLMs to generate and encode text, introducing external knowledge. However, integrating LLMs into concept recommendation presents two urgent challenges: 1) How to construct text for concepts that effectively incorporate the human knowledge system? 2) How to adapt non-smooth, anisotropic text encodings effectively for concept recommendation? 
In this paper, we propose a novel \textbf{S}tructure and \textbf{K}nowledge \textbf{A}ware \textbf{R}epresentation learning framework for concept \textbf{Rec}ommendation (\textbf{SKarREC}). We leverage factual knowledge from LLMs as well as the precedence and succession relationships between concepts obtained from the knowledge graph to construct textual representations of concepts. Furthermore, we propose a graph-based adapter to adapt anisotropic text embeddings to the concept recommendation task. This adapter is pre-trained through contrastive learning on the knowledge graph to get a smooth and structure-aware concept representation. Then, it's fine-tuned through the recommendation task, forming a text-to-knowledge-to-recommendation adaptation pipeline, which effectively constructs a structure and knowledge-aware concept representation. Our method does a better job than previous adapters in transforming text encodings for application in concept recommendation. Extensive experiments on real-world datasets demonstrate the effectiveness of the proposed approach.
\end{abstract}

\maketitle

\section{Introduction}
Online education is increasingly becoming a vital mean for individuals to acquire knowledge and skills. To serve learners' knowledge acquisition, online education platforms recommend appropriate learning content to them. This content can vary in scope, including both broad course recommendations and detailed exercise suggestions~\cite{liu2018finding, lin2021adaptive}. This article focuses on knowledge concept recommendations, which align with the users' needs to acquire specific skills or understand certain concepts when they come to the platform.

To enhance learners' mastery levels, the strategy for planning their concept learning paths should be based on two key types of information: 1) The learner's knowledge state (learner side), which refers to their proficiency in the concepts - what they have mastered well and what they haven't. The prediction of a learner's knowledge state can be achieved through knowledge tracing models, and many related models have been proposed in this area~\cite{yang2021gikt, long2022improving, liu2022pykt}. 2) The human knowledge system (concept side), which is the semantic meanings and structural relationships between concepts. Human teachers often make educational plans based on a good understanding of each concept. For instance, ``multiplication'' logically follows ``addition'' in learning progression due to its conceptual dependency on ``addition''. Many previous approaches have introduced relationships between concepts through knowledge graphs, yet these methods overlook the intrinsic semantics of the concepts. Consequently, while they capture how learners interact with concepts and the dependencies between them, they fail to integrate concrete, factual knowledge about the concepts, leaving potential deeper semantic connections between concepts unexplored.



\begin{figure}[t]
    \centering
    \includegraphics[width=\columnwidth]{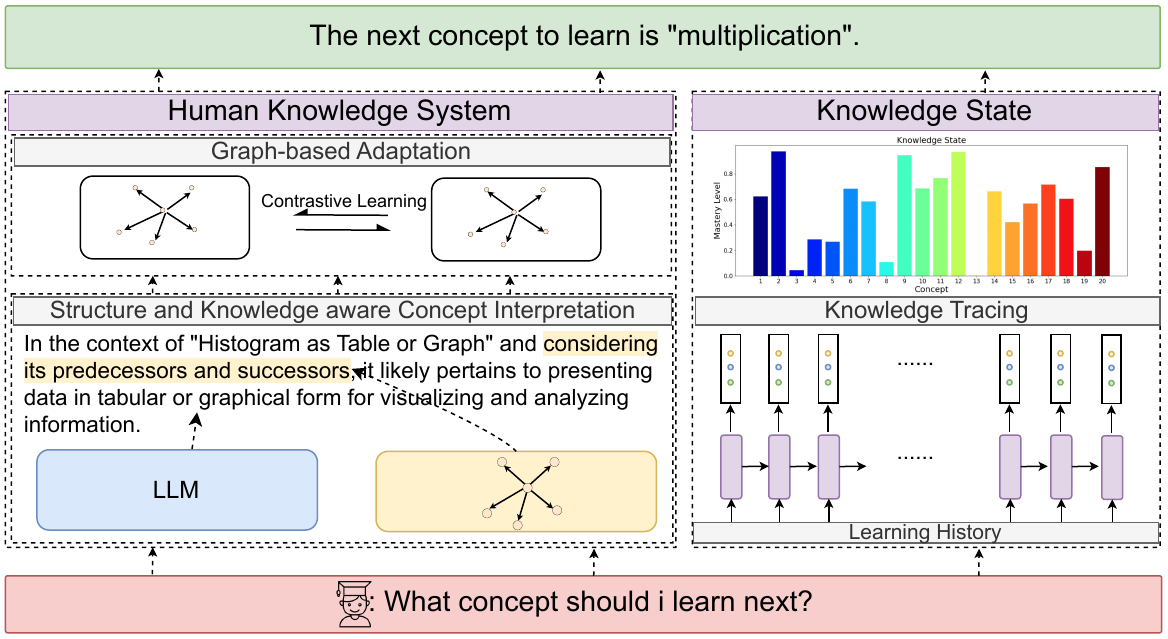}
    \centering
    \vspace{-2mm}
    \caption{The concept recommendation strategy of combining the learner's knowledge state and the human knowledge system. The knowledge state is estimated by knowledge tracing, and the human knowledge system is modeled by graph-adapted encoding of the enhanced text of each concept.}
    
    \label{fig:introFig}

\end{figure}

Nowadays, the development of Large Language Models (LLMs) provides a way of introducing the semantic meanings of the concepts. 
In product recommendation, many studies have utilized LLMs to generate and encode textual information to assist in the representations of items for recommendation purposes. However, introducing the knowledge from LLMs to represent concepts faces two challenges: \textbf{C1}) How can concepts be represented in a way that includes information from the human knowledge system? A direct way is to use the text of the concept's name, but some concepts may have ambiguous definitions in different learning contexts. For instance, ``Table'' can mean a piece of furniture in everyday life but also refers to a tabular data representation in data science. Hence, constructing appropriate textual information to represent a concept is a key issue. \textbf{C2}) How to adapt the language model's encodings of the text to the concept recommendation problem? Research on language models indicates that text encodings produced by them have non-smooth and anisotropic distributions~\cite{li2020sentence}, meaning that sentences similar in the semantic way are not similar in the embedding space, which are not suitable for direct use in downstream tasks. Therefore, previous works often establish an adapter to transfer the text encodings to suit the recommendation task. However, these adapters typically have two drawbacks: \romannumeral1) They are structurally simple, often built with one or combining multiple MLPs in a mixture of experts (MoE) way, and \romannumeral2) They lack a training target design. They are often trained end-to-end alongside the final recommendation task, lacking a tailed training objective that guarantees the effectiveness of the text adaptation. 
Therefore, using LLMs' knowledge to introduce the human knowledge system in concept recommendation requires constructing concept texts that include the semantic meanings as well as relationships between concepts and building a text adapter that transforms text representations into a space that better assists concept recommendation.

In light of the challenges, we propose \textbf{S}tructure and \textbf{K}nowledge \textbf{A}ware \textbf{R}epresentation learning framework for concept \textbf{Rec}ommen-dation (\textbf{SKarREC}). As shown in Figure ~\ref{fig:introFig}, we utilize the knowledge from the concept side (external factual knowledge provided by LLMs and relationships between concepts provided by the knowledge graph) and the learner side (the knowledge state predicted by knowledge tracing) to assist sequence models in recommending concepts. Specifically, for expressive concepts interpretation (\textbf{C1}), we develop a structure and knowledge-aware method that uses LLMs to generate concept explanation with ambiguity resolution. Combining the relationship between concepts, we construct the text description of each concept, which is fed into language models' encoders for text encoding. To effectively adapt text encodings tailed for concept recommendation (\textbf{C2}), we introduce a graph-based adapter. Specifically, the adapter is based on a graph neural network (GNN) and trained by contrastive learning in a self-supervised manner, transferring the representation from text space to the representation space that aligns with the knowledge structure (referred to as knowledge space). Thus, the final representations encompass the human knowledge system from LLMs as well as the precedence and succession relationships (structure information) of concepts in the knowledge graph. Finally, the recommendation is made based on the learner's knowledge state and the human knowledge system information contained in the concepts' representations. 

Our contributions are summarized as follow:

\begin{itemize}[topsep=5pt,leftmargin=10pt]
    \item We propose \textbf{S}tructure and \textbf{K}nowledge \textbf{A}ware \textbf{R}epresentation learning framework for concept \textbf{Rec}ommendation (\textbf{SKarREC}). By bringing open-world factual knowledge from LLMs into concepts' representation, we integrate potential deeper semantic connections for recommendation. To the best of our knowledge, we are the first to introduce large language model's knowledge into concept recommendation scenario.
    \item We utilize the structural relationships between concepts to help LLMs generate their explanation, which resolves the ambiguity problem in concept explanation.
    \item We construct a novel graph-based adapter for text encoding adaptation. The adapter is pre-trained through contrastive learning on the knowledge graph, getting a smooth and structure-aware concept representation to help with concept recommendation.
    \item We have conducted extensive experiments to show the effectiveness of our methods. Results demonstrate that the proposed approach outperforms previous baselines and improves the knowledge usage efficiency of text information.
\end{itemize}

\section{Related Work}
\subsection{Concept Recommendation}
In concept recommendation, prior research has predominantly revolved around two distinct problem formulations. The first formulation is grounded in the construction of simulators, which primarily employ simulators to emulate the progression of a learner's knowledge and utilize the simulator to evaluate the recommendation performance. Owing to the interactive nature of the simulator, reinforcement learning(RL) techniques have demonstrated remarkable effectiveness in this context, like CSEAL~\cite{liu2019exploiting}, GEHRL~\cite{li2023graph}, etc. Such methods usually take the learner's promotion from the simulator as the reward to train RL models, which naturally beat other non-RL methods.

Conversely, the focus of this paper diverges toward the second problem formulation, which concerns next-item prediction. In this setting, the primary objective of the model is to accurately align with the existing learning paths of learners as reflected in the dataset, which takes the learners' choice as the best choice instead of the simulator's prediction. In this setting, the graph-based methods mainly take advantage of using the structured information between learners and concepts. For example, ACKRec~\cite{gong2020attentional} leverages both content information and context information to learn the representation of entities via a graph convolution network. The drawback of this kind of method is that they are mostly built on heterogeneous graphs. They demand high-quality data. This is why such methods are typically tested only on comprehensive datasets like MOOCCube~\cite{yu2020mooccube}, where all necessary information is readily available. However, many platforms do not naturally possess such heterogeneous graphs, which significantly limits the applicability of these methods in various contexts. This paper explores the next concept prediction form, delving into the intricacies of aligning educational recommendations with individual student interests and choices, thereby contributing to a more personalized and effective learning experience.

\vspace{-3mm}
\subsection{LLM-enhanced Recommendation}
With the development of LLMs, more and more works have been proposed by using LLMs for recommendation~\cite{lin2023can}. LLMs have two main advantages in recommendation: 1) Interpret text information. That being said, the text information in the recommendation scenario that could not be fully used before could be utilized now. 2)Introducing open-world knowledge. LLMs are trained on a massive corpus that contains almost all the human knowledge, which makes it suitable to be used as an external knowledge base for recommendation tasks. 

By taking the two advantages, LLMs are used in different ways. One way is to use the open-world knowledge and reasoning ability of the LLMs. That being said, this kind of method is mainly based on prompt engineering to make the LLMs generate the recommendation-related text information they want. For example, P5~\cite{geng2022recommendation} presents a text-to-text paradigm for recommendation. It converts all data into natural language sequences and trains the language model with a language modeling task. In this approach, the recommendation is taken as a natural language processing problem and depends more on the LLMs' inherent reasoning ability learned from the pre-training corpus.
The other kind is to use it to generate dense vectors as some sort of representation or feature. For example, UniSRec~\cite{hou2022towards} encodes each item using a language model and develops a mixture-of-experts (MoE) adaptor that transforms the text semantic into a universal form suited to the cross-domain recommendation tasks. RECFORMER ~\cite{li2023text}, on the other hand, re-train Longformer~\cite{beltagy2020longformer} model on the recommendation datasets with mask-item-prediction and mask-token-prediction tasks, which develops a recommender system that is purely based on text information. Despite all the success these methods get, they paid little attention to the adapter that adapts the text representation to the recommendation task, which could cause the text information not to be fully utilized.

\vspace{-3mm}
\section{Preliminaries}
In this section, we formulate the task of concept recommendation and the knowledge graph.

\textbf{Concept Recommendation Task. }In the concept recommendation scenario, we have a set of learners, denoted by $\mathcal{U}$, a set of concepts denoted by $\mathcal{K}$. At time step $t$, we recommend a concept $k\in \mathcal{K}$ based on the learner's learning history $\mathcal{H}_t$. The history is constructed as a sequence of tuples $\mathcal{H}_t=((k_1, c_1), (k_2,c_2)...(k_t, c_t))$, where $k_t$ is the concept learned at time step $t$; $c_t \in \{0, 1\}$ representing the feedback of the learner (like the answer correctness of the concept-related question). Our goal is to recommend $k_{t+1}$ based on $\mathcal{H}_t$. 

Each concept $k \in \mathcal{K}$ corresponds to a text corpus $w^{k}$, like ``addition'' or ``venn diagram'' that could be used as auxiliary information for the recommendation. 

\textbf{Knowledge Graph.} The concepts of a certain area could be formulated as a graph representing the prerequisite relationships between them. The knowledge graph is a directed graph $G=\{\mathcal{K}, \mathcal{E}\}$, where each node $k$ is a concept, and each edge $e$ from $k_1$ to $k_2$ represents that concept $k_1$ is the prerequisite concept of $k_2$. However, since such graphs are often not readily available in many situations or datasets, we construct transition graphs\cite{nakagawa2019graph} to approximate these knowledge graphs. In these transition graphs, an edge $e$ from $k_1$ to $k_2$ is drawn if, typically, $k_2$ is learned after $k_1$ according to the sequences observed in learners' activities. If the graph is provided, then we use it directly.
\section{Methodology}

\begin{figure*}[t]
    \centering
    \includegraphics[width=1.0\linewidth]{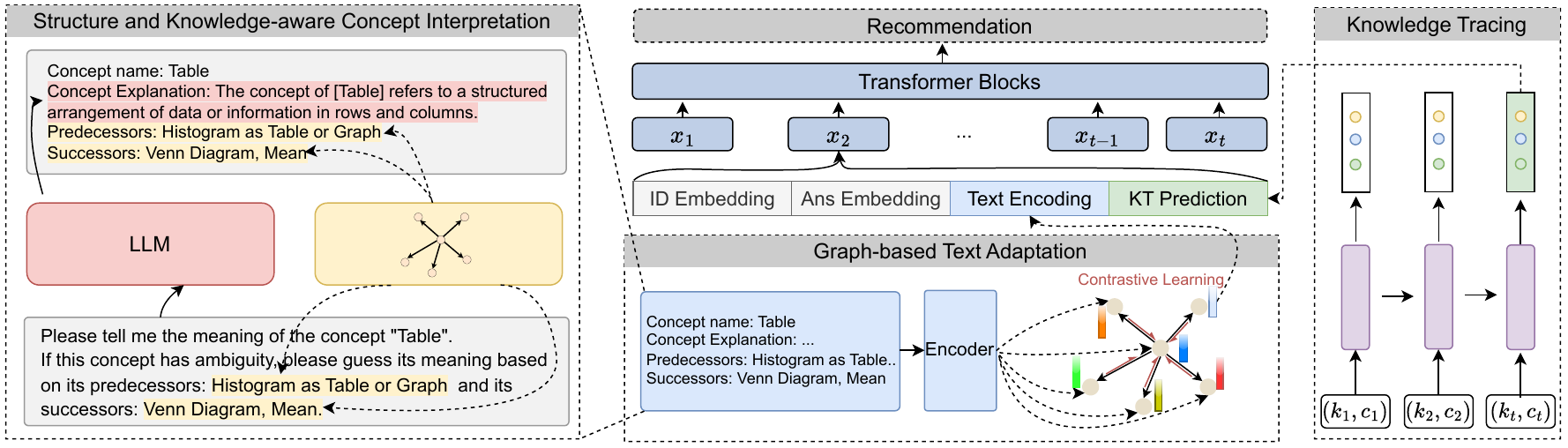}
    \vspace{-2mm}
    \caption{The overall framework of \our. The left part shows the construction of concept interpretations integrates structure and knowledge. The right part is the demonstration of knowledge tracing. The bottom center details the graph-based text adaptation process. The top middle outlines the recommendation mechanism using transformed embeddings. }
    \vspace{-2mm}
    \label{fig:overall}
\end{figure*}

\subsection{Overview}
We have developed a concept recommendation framework that integrates the knowledge states of learners and the human knowledge system, as shown in Figure ~\ref{fig:overall}. To incorporate the human knowledge system using LLMs and knowledge graphs and to adapt this information into text for the recommendation process, we introduced three modules:

\textbf{Structure and Knowledge-aware Concept Interpretation.} 
The human knowledge system is based on the definitions and relationships of concepts. To incorporate them into the recommendation process, we enhance the textual descriptions of concepts through LLMs and the knowledge graph. By using the knowledge graph, we enable the LLM to generate unambiguous semantic interpretations of each concept. Subsequently, we present the name, explanation, and preceding and succeeding nodes of the concepts as the concept's text description for the subsequent recommendation.

\textbf{Graph-based Text Adaptation.} To improve how we transform text into recommendations, we develop a graph-based adapter. This adapter adopts contrastive learning on the graph to optimize the adaptation process, which emphasizes the structural connections between concepts, enhancing the effectiveness of the final recommendation task.

\textbf{Learner Knowledge State Representation.} In addition to modeling the concept side by incorporating the human knowledge system, we have introduced the results of knowledge tracing as auxiliary information on the learner side. This approach ensures that the concept recommendation not only considers the relevance between concepts but also bases recommendations on the current knowledge state of the learner.

Next, we go through the details of the framework.

\vspace{-3mm}
\subsection{Structure and Knowledge-aware Concept Interpretation}
The key feature of the concept recommendation problem is that the learner's choice is largely based on the human knowledge system, which is constructed by the semantic meanings or definitions and the relationships between concepts. Human teachers often make educational plans based on a good understanding of each concept. Therefore, incorporating the semantics of each concept is crucial for developing a concept recommendation system.

Directly using the concept name or letting a dialogue model, like GPT-3.5~\cite{achiam2023gpt}, to generate explanations for each concept would cause the ambiguity problem. That is,  a concept can have different meanings based on the field of study or the context. For instance, ``Table'' could refer to a data structure in data science or to a piece of furniture in everyday conversation.
 If the LLM is asked to explain ``Table'', it might incorrectly interpret it as a piece of furniture. To eliminate such ambiguities, we refine our prompts to the LLM by including the concept's predecessor and successor nodes of the concept from the knowledge graph. This method helps the LLM to deliver an explanation tailored to the specific educational context, greatly diminishing the risk of ambiguity.

To ensure that the text of a concept contains sufficient information, we have designed the textual representation of each concept to include the following four parts:
\begin{itemize}[topsep = 3pt,leftmargin =10pt]
    \item Concept Name: The name of the concept $w^{k}$.
    \item Concept Explanation: We utilize GPT-3.5 to infer the explanation of the concept based on its predecessors and successors on the knowledge graph.
    \item Predecessors: The names of the prerequisite concepts of this concept in the knowledge graph.
    \item Successors: The names of the successor concepts of this concept in the knowledge graph.
\end{itemize}
Each item is represented in the key-value text containing the four parts above. We denote the enhanced text as $\tilde{w}^{k}$.

\subsection{Graph-based Text Adaptation}
Due to the anisotropy nature of vanilla text embeddings~\cite{li2020sentence}, recommendation tasks often require an adapter to transfer the text embeddings to suit the task~\cite{hou2022towards}. Previously, this transfer process was trivial in two aspects: \romannumeral1) The construction of the adapter is often one or several Multi-Layer Perceptron (MLP) units. \romannumeral2) The training objective for the adapter is not specifically tailored but rather is trained in conjunction with the recommendation task, which didn't show much adaptation effect.

In educational contexts, textual information is closely linked with the human knowledge system. Our idea is to adapt educational texts based on the structure of human knowledge rather than merely using a simple MLP and training it alongside recommendation tasks. In light of this, we propose a graph-based adapter that is built upon a GNN and trained by contrastive learning on the graph.

\subsubsection{Concept Text Encoding} 
After gaining the enhanced text of each concept $\tilde{w}^{k}$, we encode the text utilizing the pre-trained language model BART~\cite{lewis2019bart} due to its straightforward architecture and considerable text comprehension capabilities. Given a concept $k$ and  its corresponding enhanced text $\tilde{w}^{k}=\{ \tilde{w}^{k}_1, \tilde{w}^{k}_2, ... \tilde{w}^{k}_c \}$. We add a special token [CLS] at the beginning and use the language model to encode.
\begin{equation}
    T^{k}=LM([[CLS], \tilde{w}^{k}_1, \tilde{w}^{k}_2, ... \tilde{w}^{k}_c])
\end{equation}
where $T^{k}$ is the final hidden vector corresponding to token [CLS] as the encoding of the enhanced text of concept $k$; $c$ is the length of the word sequence $\tilde{w}^{k}$.

\subsubsection{Graph-based Adapter}
To get a smooth and structure-aware concept representation, we propose to learn from the knowledge graph $G=\{\mathcal{K}, \mathcal{E}\}$. Specifically, we first set the initial encoding of each node as the text encoding of the corresponding concept.
\begin{equation}
    h^{0}_{k} = T^{k}
\end{equation}
Then we utilize Graph Convolutional Network (GCN)~\cite{kipf2016semi} to learn the embedding of each concept for its simplicity and effectiveness. 
\begin{equation}
    h^{l}_{k} = Combine(H(h^{l-1}_{k}), Agg({H(h^{l-1}_{i})|i\in \mathcal{N}_{k}}))
\end{equation}
where $h^{l}_{k}$ denotes the node representation of concept $k$ at the l-th layer; $h^{l-1}_{k}$ is that of previous layer; $H(\cdot)$ is a non-linear transformation function (usually implemented by MLP);  $Agg(\cdot)$ is the aggregation function of the neighbors' embeddings of node $k$;  $Combine(\cdot)$ denotes the combination function of the previous layer's representations of node $k$ and the aggregation result. 

Using a GNN as the adapter for concept recommendation offers two key benefits: 1) In terms of the distribution of embeddings, GNN's learning mechanism can result in smoother embeddings~\cite{li2018deeper}, potentially resolving the problem of anisotropy in text embedding distributions. 2) Regarding the concept recommendation task, GNN's ability to aggregate information allows each concept's embedding to incorporate relationships between concepts, leading to recommendations that are more aligned with the underlying prerequisite relationships between concepts.

\subsubsection{Graph-based Learning}
The aim of utilizing a knowledge graph is to enhance the suitability of vanilla text embeddings for concept recommendation tasks. We achieve this by transforming the embedding of concepts from the text representation space to a representation space that aligns with the knowledge structure (referred to as knowledge space). This is done through learning on the graph. To accomplish this, we employ the contrastive learning method to train the graph adapter. It trains the adapter to learn stable node representations despite perturbations in the graph structure, thereby capturing the graph's deep structural information.

Specifically, we first gain different views of the knowledge graph by edge dropout~\cite{wu2021self}:
\begin{equation}
    v_1(G) = (\mathcal{K}, \mathbf{M}_1 \odot \mathcal{E}), v_2(G) = (\mathcal{K}, \mathbf{M}_2 \odot \mathcal{E})
\end{equation}
where $\mathbf{M}_1, \mathbf{M}_2 \in \{0, 1\}^{|\mathcal{E}|}$ are two masking vectors on the edge set. A hyper-parameter $\gamma$ is set to be the masking ratio. 

By randomly masking the graph edges by a certain ratio, we get two sets of edges corresponding to two views of the knowledge graph. We treat the identical nodes across different views as positive pairs and different nodes from these views as negative pairs. We then train the adapter through InfoNCE loss~\cite{oord2018representation}: 
\begin{equation}
    \mathcal{L}_{ssl}^{G} = -\sum_{k\in \mathcal{K}}log\frac{e^{sim(h_k^{(1)}, h_k^{(2)})/\tau}}{\sum_{i\in \mathcal{K}, i\neq k}e^{sim(h_k^{(1)}, h_{i}^{(2)})/\tau}}
\end{equation}
where $sim(\cdot)$ is the similarity measurement function, which is set as the cosine similarity; $h^{(1)}$ and $h^{(2)}$ are two representation of the nodes from two different views; $\tau$ is a temperature parameter.

The goal of introducing contrastive learning on the knowledge graph is to enhance the vanilla text embeddings and transfer them into a revamped embedding space. Within this space, the representation of each concept is enriched with relational characteristics from the knowledge graph, ensuring that nodes structurally akin to each other also exhibit similarity in the embedding space.

\subsection{Learner Knowledge State Representation}
Personalized concept recommendation should not only align with the human knowledge system but also be based on the individual learner's knowledge state. Therefore, we employ the widely used deep knowledge tracing (DKT) model ~\cite{piech2015deep} to track the learner's knowledge state. The learning history $\mathcal{H}_t$ is encoded by a recurrent neural network (RNN) to predict the learner's mastery level of all concepts. We choose the gated recurrent unit (GRU)~\cite{chung2014empirical} network as the sequence encoder due to its simplicity and effectiveness.
\begin{equation}
    s_t = GRU(\mathcal{H}_t)
\end{equation}
The training objective of DKT is the prediction of the correctness $c_{t+1}$ of the next concept $k_{t+1}$:
\begin{equation}
    \mathcal{L}^{KT}=(s_t(k_{t+1}) - c_{t+1})^2
\end{equation}
The knowledge state of the learner is consistently changing, so we predict the knowledge state representation at each time step and concatenate it to the input of the recommendation model.

\subsection{Self-Attention Based Recommendation}
Having the encoding of the graph-adapted textual representation of the concept $h_{k_t}$ and the knowledge state $s_{t}$, we further add two embedding layers to map the IDs of concepts and the answers. Specifically, we have a concept embedding matrix $M_{K} \in \mathbb{R}^{|\mathcal{K}|\times d}$ and an answer embedding matrix $M_{A} \in \mathbb{R}^{2 \times d}$ representing answering right or wrong. Each time step's record $(k_t, c_t)$ is encoded by four parts of embeddings:
\begin{equation}
    x_{t} = i_{k_t} \oplus a_{c_t} \oplus  h_{k_t} \oplus  s_{t}
\end{equation}
where $x_{t}$ is the final representation; $i_{k_t} \in M_{K}$ is the id embedding of $k_t$; $a_{c_t} \in M_{A}$ is the answer embedding of $c_t$; $h_{k_t}$ is the graph-adapted encoding of the concept; $s_{t}$ is the knowledge state representation.

Given a sequence of the encoding of the learning record $(x_1, x_2, x_3... x_t)$ , we further utilize the widely used Transformer structure~\cite{vaswani2017attention} to encode the sequence and 
\begin{equation}
    [F_1, F_2, ... F_t]=FFN(Atten([x_1, x_2, ...x_t]))
\end{equation}
where $FFN(\cdot)$ is the feed forward network; $F_t\in \mathbb{R}^{d}$ is the output encoding at time step $t$; $Atten(\cdot)$ is the Multi-head attention algorithm.

We calculate the scores for all concepts:
\begin{equation}
    P(\cdot|\mathcal{H}_{t})=W^{T} F_{t}
\end{equation}
where $W^{T}\in \mathbb{R}^{|\mathcal{K}|\times d}$ is a learnable parameter matrix.

\subsection{Overall Training Procedure}
Here we present our pre-training and fine-tuning procedure, the detailed algorithm could be found in the Appendix ~\ref{sec:algo}  Algorithm ~\ref{algo}.
\subsubsection{Pre-training}
During the pre-training phase, our training focuses on three parts: 1) Graph-based pre-training. Train the graph adapter to transfer embeddings from text space into a graph or knowledge space. 2) Knowledge-Tracing pre-training. Train the knowledge tracing module to represent a learner's knowledge state accurately. 3)Sequence-based pre-training. Train the Transformer network to utilize the graph-adapted encodings and knowledge tracing encodings to understand the sequential order of learning sequences. The former two could be achieved by the contrastive learning loss on the knowledge graph $\mathcal{L}_{ssl}^{G}$ and the knowledge tracing loss $\mathcal{L}^{KT}$, respectively. 

For the sequence-based pre-training, various training strategies have been suggested, all revolving around the principle of partially hiding sequence information and then predicting it through sequence modeling. Here, we adopt the same scheme as S3Rec~\cite{zhou2020s3} since it covers most of the sequence-based self-supervised learning objectives: mask item prediction ($\mathcal{L}_{MIP}$), mask segment prediction ($\mathcal{L}_{MSP}$), mask attribute prediction ($\mathcal{L}_{MAP}$), and associated attribute prediction ($\mathcal{L}_{AAP}$). The final sequence-based pre-training objective is:
\begin{equation}
    \mathcal{L}_{ssl}^{seq} = \mathcal{L}_{MIP} + \mathcal{L}_{MSP} + \mathcal{L}_{MAP} + \mathcal{L}_{AAP}
\end{equation}
In summary, in the pre-training stage, we pre-train graph-based adapter, knowledge tracing module and Transformer network by $\mathcal{L}_{ssl}^{G}$, $\mathcal{L}^{KT}$ and $\mathcal{L}_{ssl}^{seq}$.

\subsubsection{Fine-tuning}
After pre-training, we fine-tune the whole model in an end-to-end manner based on the concept recommendation task. We adopt the cross-entropy loss:
\begin{equation}
    \mathcal{L}_{rec} = -\frac{1}{N}\sum_{n=0}^{N}\sum_{j=0}^{|\mathcal{K}|}y_{nj}log(P_{nj})
\end{equation}
where $N$ is the training sample number; $y_{nj}$ is a binary indicator of whether concept $j$ is the actual next one of sample $n$; $P_{nj}$ is the predicted probability of the concept $j$ of sample $n$.

\begin{table}
  \caption{Dataset Statistics}
  \label{tab-datasetsta}
  \begin{tabular}{lccc}
    \toprule
      Dataset   & Junyi & ASSIST12 & ASSIT09\\
    \hline
    \#Concepts & 835 &  265 & 145 \\
    \#Learners & 341,195  & 24,155 & 3,322  \\
    \#Records &21,460,249 & 1,853,338 & 187,914  \\
    Correct  rate& 54.38\% & 69.55\%  & 64.12\% \\
    \bottomrule
  \end{tabular}
\end{table}

\begin{table*}[t]
\centering
\small
\caption{Performance comparison of different recommendation models. The best and the second-best performance is bold and underlined respectively.``*'' denotes that the improvement are significant at level of $p<0.05$ with paired t-test comparing with the second best baseline.}

\label{tab:overallperf}
\scalebox{1.0}{
\setlength{\tabcolsep}{1mm}{
\begin{tabular}{llcccccccccc}
\toprule

\multicolumn{1}{c}{\textbf{}}  & \multicolumn{1}{c}{\textbf{}}  & \multicolumn{2}{c}{\textbf{Graph-based Methods}}  & \multicolumn{2}{c}{\textbf{ID-Only Methods}}   & \multicolumn{2}{c}{\textbf{Text-Only Methods}} & \multicolumn{4}{c}{\textbf{ID-Text Methods}}   \\

\cmidrule(lr){3-4} \cmidrule(lr){5-6} \cmidrule(lr){7-8} \cmidrule(lr){9-12}

\textbf{Dataset} & \textbf{Metric} & \textbf{ACKRec} & \textbf{GCARec} & \textbf{SASRec} & \textbf{BERT4Rec}   & \textbf{ZESRec}       & \textbf{RECFORMER} & \textbf{FDSA} & \textbf{S3-Rec} & \textbf{UniSRec} & \textbf{\our}  \\ 

\midrule

\multirow{3}{*}{ASSIST09}         
& HR@1    & 0.2408  & 0.2513   & 0.7649   & 0.1935  & 0.5809   & 0.6072     & 0.7627  & \underline{0.7661}   & 0.7597  & \textbf{0.7922$^{*}$}     \\
& NDCG@5  & 0.3787  & 0.3904   & \underline{0.8736}    & 0.2203  & 0.7269   & 0.6540     & 0.8722  & 0.8731   & 0.8647  & \textbf{0.8838$^{*}$}    \\
& MRR     & 0.3627  & 0.3756    & 0.8503   & 0.2342  & 0.6998   & 0.6711     & 0.8484  & \underline{0.8505}   & 0.8439  & \textbf{0.8646$^{*}$}    \\ 

\midrule

\multirow{3}{*}{ASSIST12}         
& HR@1    & 0.1976  & 0.2119  & 0.2838   & 0.2540  & 0.2596   & 0.1990    & 0.2861  & \underline{0.2904}   & 0.2888  & \textbf{0.2933$^{*}$}    \\
& NDCG@5  & 0.3078  & 0.3169  & 0.4895    & 0.4247  & 0.4649   & 0.2055     & 0.4886  & \textbf{0.4957}   & 0.4934  & \underline{0.4954}  \\
& MRR     & 0.3064  & 0.3182   & 0.4574   & 0.4047  & 0.4343   & 0.2536     & 0.4579  & \underline{0.4633}   & 0.4617  & \textbf{0.4645$^{*}$}     \\ 
\midrule

\multirow{3}{*}{Junyi}                
& HR@1    & 0.1447  & 0.1284  & 0.8469   & 0.3550  & 0.8546   & 0.8279     & 0.8492  & \underline{0.8608}   & 0.8407  & \textbf{0.8730$^{*}$}    \\
& NDCG@5  & 0.1872  & 0.1650  & 0.8907    & 0.4826  & 0.8930   & 0.7625     & 0.8917  & \underline{0.9022}   & 0.8890  & \textbf{0.9076$^{*}$}   \\
& MRR     & 0.1918  & 0.1740    & 0.8825   & 0.4661  & 0.8860   & 0.8497     & 0.8840  & \underline{0.8942}  & 0.8799  & \textbf{0.9014$^{*}$}    \\ 

\bottomrule
 
\end{tabular}
}}
\end{table*}
\section{Experiments}
\subsection{Experimental Settings.}
\subsubsection{Datasets.}
We conducted experiments through three datasets: \textit{Junyi\footnote{https://www.kaggle.com/datasets/junyiacademy/learning-activity-public-dataset-by-junyi-academy}},  \textit{ASSIT12\footnote{https://sites.google.com/site/assistmentsdata/datasets/2012-13-school-data-with-affect}} and \textit{ASSIT09\footnote{https://sites.google.com/site/assistmentsdata/home/2009-2010-assistment-data}}. They all provide learning sequences of learners. These datasets are transformed from their original question-answer pair sequences into concept-answer pairs. Each concept has an ID and a name. We provide the statistics of these three datasets in Table ~\ref{tab-datasetsta}.

For the knowledge graph, \textit{Junyi} dataset offers a dedicated knowledge graph that outlines concept dependencies. While for \textit{ASSIST09} and \textit{ASSIST12}, we utilize transition graphs\cite{nakagawa2019graph} as an estimation of knowledge graphs.

\subsubsection{Baselines.}
We are comparing four different sets of baseline methods. These include methods that use only concept IDs, methods that utilize item IDs and treat the text description of items as additional information, methods that rely solely on the text of the concepts as input, and methods based on heterogeneous graphs proposed for concept recommendation.

(1) ID-Only methods:
\begin{itemize}[topsep = 3pt,leftmargin =15pt]
    \item \textbf{SASRec}~\cite{kang2018self}: utilizes a self-attention model to predict the next item.
    \item \textbf{BERT4Rec}~\cite{sun2019bert4rec}: utilizes the original Bert~\cite{devlin2018bert} scheme for next item prediction.
\end{itemize}

(2) ID-Text methods:
\begin{itemize}[topsep = 3pt,leftmargin =15pt]
    \item \textbf{FDSA}~\cite{zhang2019feature}: utilizes self-attention to capture item and feature transition patterns.
    \item \textbf{S3-Rec}~\cite{zhou2020s3}: pre-trains self-attention models with mutual information maximization objectives related to attributes, items, sub-sequences, etc.
    \item \textbf{UniSRec(Transductive)}~\cite{hou2022towards}: uses an MoE-based adapter to transit the information from text to recommendation area. We choose to compare with the model fine-tuned in transductive setting using both ID and text considering that it gets better results than text-only setting in the original paper.
\end{itemize}

(3) Text-Only methods:
\begin{itemize}[topsep = 3pt,leftmargin =15pt]
    \item \textbf{ZESRec}~\cite{ding2021zero}: utilizes a pre-trained language model to encode the item texts as the input feature for next-item prediction.
    \item \textbf{RECFORMER}~\cite{li2023text}: threats item key-value attributes as texts and formulates the interaction sequences as sentences sequences. Then a pre-trained language model is fine-tuned for the next item prediction.
\end{itemize}

(4) Graph-based concept recommendation methods:
\begin{itemize}[topsep = 3pt,leftmargin =15pt]
    \item \textbf{ACKRec}~\cite{gong2020attentional}: builds a heterogeneous graph an uses attention mechanism to learn the representation from different meta-paths.
    \item \textbf{GCARec}~\cite{yu2023graph}: develops topology-level and feature-level augmentations to generate different views of learner-concept graph to conduct contrastive learning.
\end{itemize}

\subsubsection{Evaluation metrics.}
To evaluate the performance of sequential recommendation, we adopt Hit Ratio (HR), Normalized Discounted Cumulative Gain (NDCG), and Mean Reciprocal Rank (MRR) as the evaluation metrics.
We employ a leave-one-out strategy, where each sequence's last concept is reserved for testing, the second to last for validation, and all preceding ones for training.

\subsection{Overall Performance}

Table ~\ref{tab:overallperf} reports the overall performance of all methods in three datasets. It demonstrates that:
\begin{itemize}[topsep = 3pt,leftmargin =10pt]
    \item Our proposed \our achieves the best overall performance. Different from baselines, our proposed enhanced text and graph-based adapter make the recommendation model understand the structure and semantic links between concepts, leading to improved performance in concept recommendation.
    \item Adding textual information to concept recommendations proves beneficial, allowing methods that use both IDs and text (ID-Text methods) to outperform ID-Only methods. Nonetheless, the improvement is constrained by the absence of a dedicated adapter in earlier approaches. With the introduction of the graph-based adapter, the advantages of incorporating text into concept recommendations are more significantly realized.
    \item The performance of graph-based methods designed for concept recommendation declines in the absence of heterogeneous information. Originally, these methods relied on heterogeneous graphs that included a variety of elements like courses, videos, and teachers. However, such comprehensive data is rarely available in most open-source educational datasets, leading to limitations in these approaches.
\end{itemize}

\subsection{Ablation Study}
We conduct ablation experiments to analyze how our proposed modules influence the final concept recommendation performance.
\subsubsection{Graph-based Adapter.}
We compare different versions of our model to assess the impact of our suggested graph-based adaptation strategy. The results are shown in Table ~\ref{tab:AbOnGraph}. In ``\our-No-GraphSSL'', we remove the contrastive learning (retaining the GNN structure, training only through the fine-tune recommendation task). In ``\our-MoE'', we entirely replaced the graph-based adapter with the MoE adapter from previous work~\cite{hou2022towards}. Both variations exhibit a notable drop in performance compared to the original structure. This not only demonstrates the graph-based adapter's superior ability to adapt the text encodings but also indicates that this capability largely stems from the contrastive learning on the graph rather than merely changing the adapter's structure to a GNN. Graph-based contrastive learning has a clear training objective: to ensure that each concept's embedding captures the prerequisite relationships between concepts. This relationship is crucial for incorporating the semantics of concepts and effectively integrating the human knowledge system, thereby enhancing recommendation outcomes.

\begin{table}[t]
\centering
\small
\caption{Ablation study about the proposed graph-based adapter. The ``-MoE'' suffix means that we change the adapter to MoE adapter and the ``-No-GraphSSL'' means that we remove the graph-based contrasrive learning.}
\vspace{-3mm}
\label{tab:AbOnGraph}
\scalebox{0.95}{
\setlength{\tabcolsep}{1mm}{
\begin{tabular}{llccc}
\toprule

\multicolumn{1}{c}{\textbf{Dataset}}  & \multicolumn{1}{c}{\textbf{Metric}}  & \multicolumn{1}{c}{\textbf{\our-No-GraphSSL}}  & \multicolumn{1}{c}{\textbf{\our-MoE}}   & \multicolumn{1}{c}{\textbf{\our}} \\

\midrule

\multirow{3}{*}{ASSIST09}         
& HR@1    & 0.7727  & 0.7754     & \textbf{0.7922}     \\
& NDCG@5  & 0.8764  & 0.8760     & \textbf{0.8838}     \\
& MRR    & 0.8543  & 0.8548     & \textbf{0.8646}   \\ 

\midrule

\multirow{3}{*}{ASSIST12}         
& HR@1    & 0.2901  & 0.2855     & \textbf{0.2933}     \\
& NDCG@5  & 0.4946  & 0.4923     & \textbf{0.4954}     \\
& MRR    & 0.4626  & 0.4601     & \textbf{0.4645}   \\ 
\midrule

\multirow{3}{*}{Junyi}                
& HR@1    & 0.8711  & 0.8717     & \textbf{0.8730}     \\
& NDCG@5  & 0.9060  & 0.9071     & \textbf{0.9076}     \\
& MRR    & 0.8998  & 0.9006     & \textbf{0.9014}   \\ 
\bottomrule

\vspace{-4mm}
\end{tabular}
}}
\end{table}

\subsubsection{Structure and Knowledge-aware Concept Interpretation.}
We now delved into the effects of the proposed structure and knowledge-aware concept interpretation, referred to as enhanced text. We conduct comparison experiments, and the results are shown in Table ~\ref{tab:AbOnText}. In the variant ``\our-No-LLM'', we drop the concept explanations provided by the LLM and only keep the information about the concept's predecessors and successors from the graph. We observe a decline in performance. It demonstrates that capturing the semantic meanings of concepts are important for introducing the human knowledge system. In the variant ``\our-Name'', we drop all the enhanced text (including LLM's Concept Explanation and the predecessors and successors from the graph) and only provide the concept names. This substitution results in a significant drop in recommendation performance, illustrating that the text names are insufficient for incorporating concepts' semantic meanings. Our proposed method successfully incorporates semantic and structural information into the text, improving the learning for concept representation.

\begin{table}[t]
\centering
\small
\caption{Ablation study about the structure and knowledge-aware concept interpretation. The ``-No-LLM'' suffix means that we drop the LLM's concept explanation and the ``-Name'' means that we drop all the enhanced text.}
\vspace{-3mm}
\label{tab:AbOnText}
\scalebox{0.95}{
\setlength{\tabcolsep}{1mm}{
\begin{tabular}{llccc}
\toprule

\multicolumn{1}{c}{\textbf{Dataset}}  & \multicolumn{1}{c}{\textbf{Metric}}  & \multicolumn{1}{c}{\textbf{\our-No-LLM}}  & \multicolumn{1}{c}{\textbf{\our-Name}}   & \multicolumn{1}{c}{\textbf{\our}} \\

\midrule

\multirow{3}{*}{ASSIST09}         
& HR@1    & 0.7898  & 0.7775     & \textbf{0.7922}     \\
& NDCG@5  & 0.8832  & 0.8756     & \textbf{0.8838}     \\
& MRR    & 0.8633  & 0.8552     & \textbf{0.8646}   \\ 

\midrule

\multirow{3}{*}{ASSIST12}         
& HR@1    & 0.2886  & 0.4926     & \textbf{0.2933}     \\
& NDCG@5  & 0.4936  & 0.4612     & \textbf{0.4954}     \\
& MRR    & 0.4617  & 0.8721     & \textbf{0.4645}   \\ 
\midrule

\multirow{3}{*}{Junyi}                
& HR@1    & 0.8708  & 0.8721     & \textbf{0.8730}     \\
& NDCG@5  & 0.9065  & 0.9069     & \textbf{0.9076}     \\
& MRR    & 0.9001  & 0.9006     & \textbf{0.9014}   \\ 
\bottomrule
 
\end{tabular}
}}
\end{table}

\begin{figure}[htbp] 
    \centering
    \begin{subfigure}[t]{0.48\columnwidth}
           \centering
           \includegraphics[width=\columnwidth]{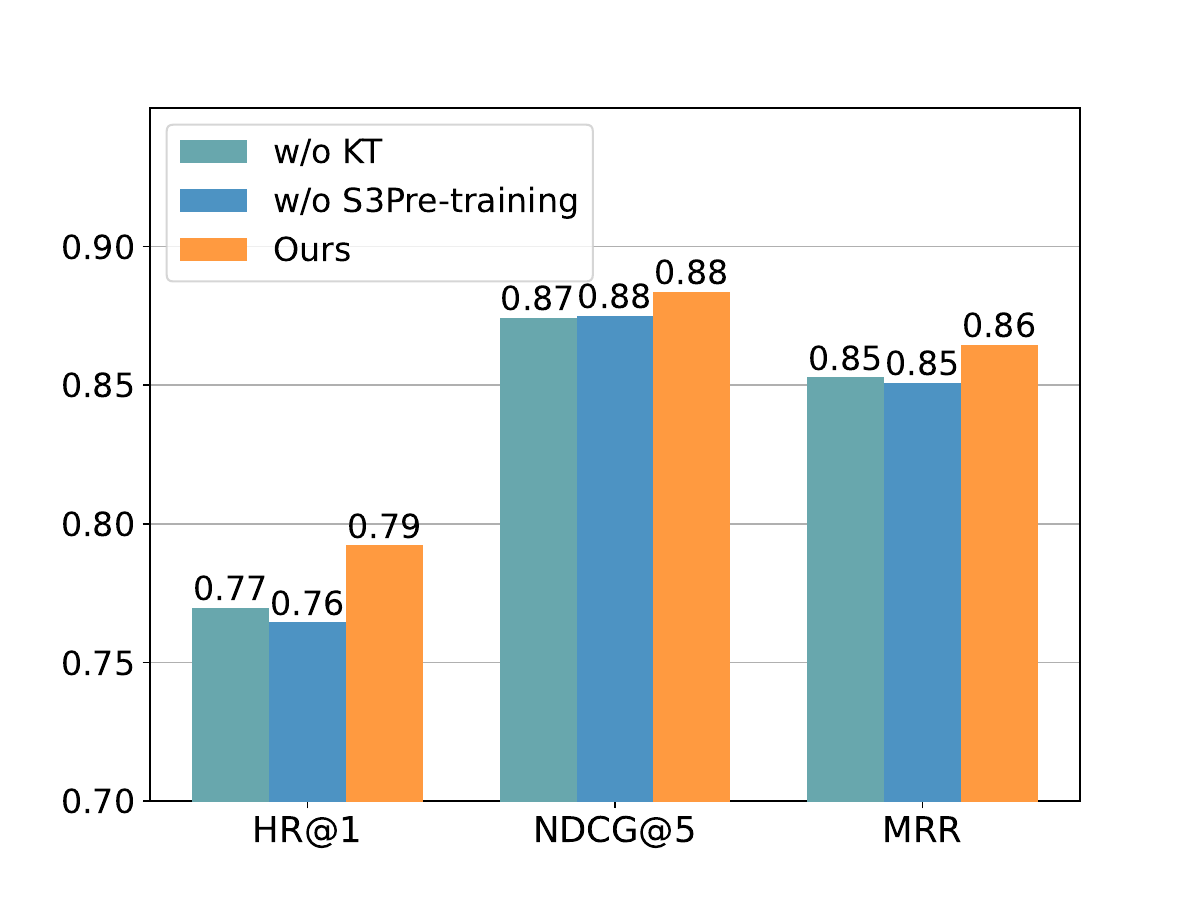}
            \caption{ASSIST09}
    \end{subfigure}
    \begin{subfigure}[t]{0.48\columnwidth}
           \centering
           \includegraphics[width=\columnwidth]{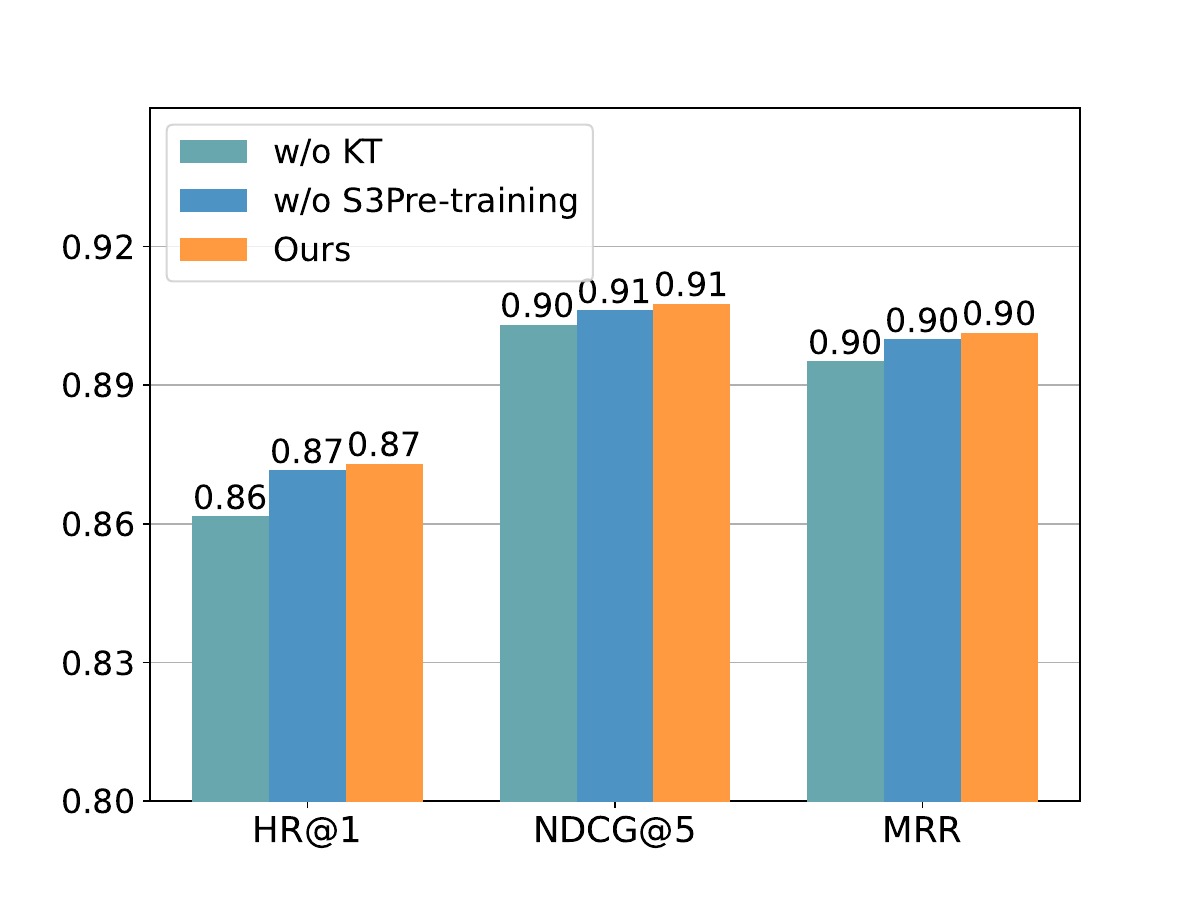}
            \caption{Junyi}
    \end{subfigure}
    \centering
    \vspace{-3mm}
    \caption{Performance of comparison with/without sequence modeling tasks.}
    \vspace{-3mm}
    \label{fig:AbOnSeq}
\end{figure}

\begin{figure*}[htbp]
    \centering
    \begin{subfigure}[t]{0.325\textwidth}
           \centering
           \includegraphics[width=\columnwidth]{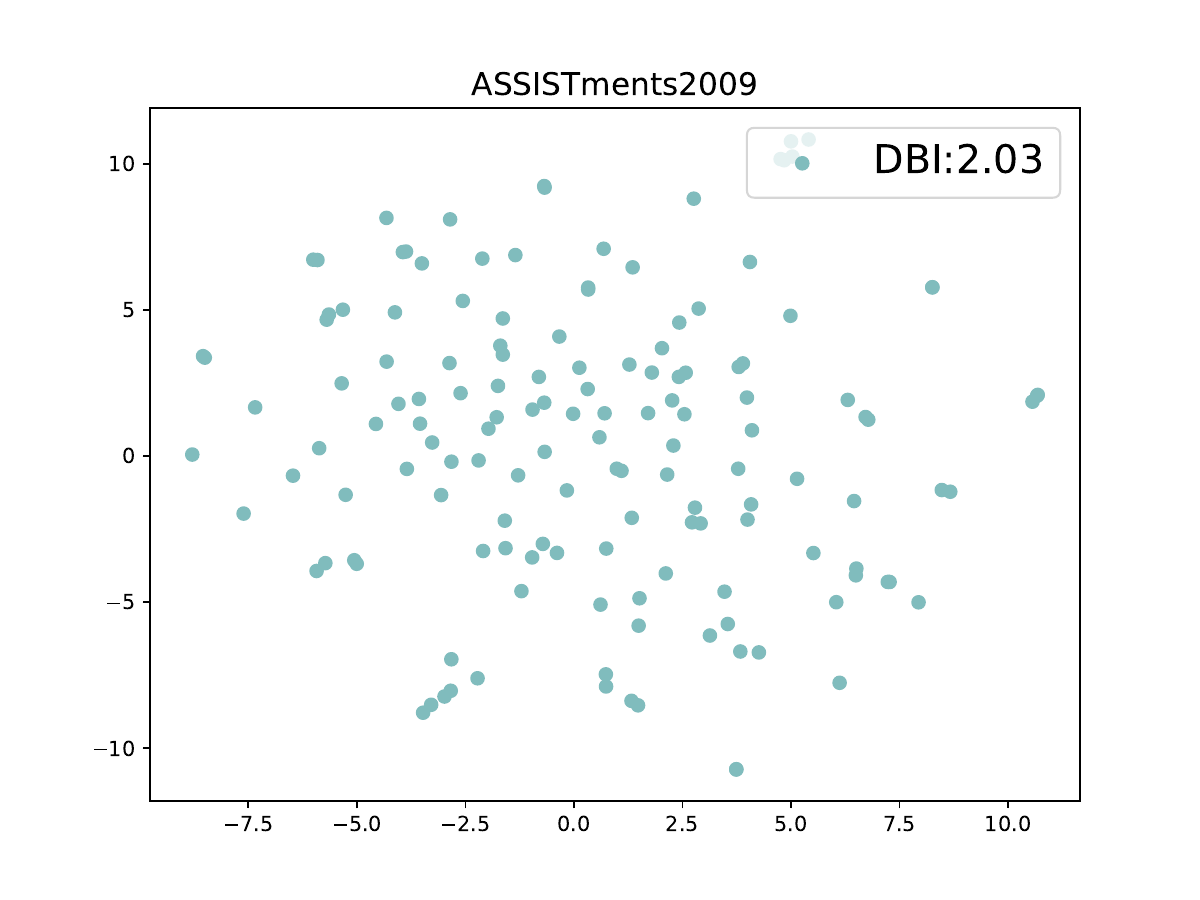}
           \vspace{-2mm}
            \caption{Vanilla Text embeddings}
    \end{subfigure}
    \begin{subfigure}[t]{0.325\textwidth}
           \centering
           \includegraphics[width=\columnwidth]{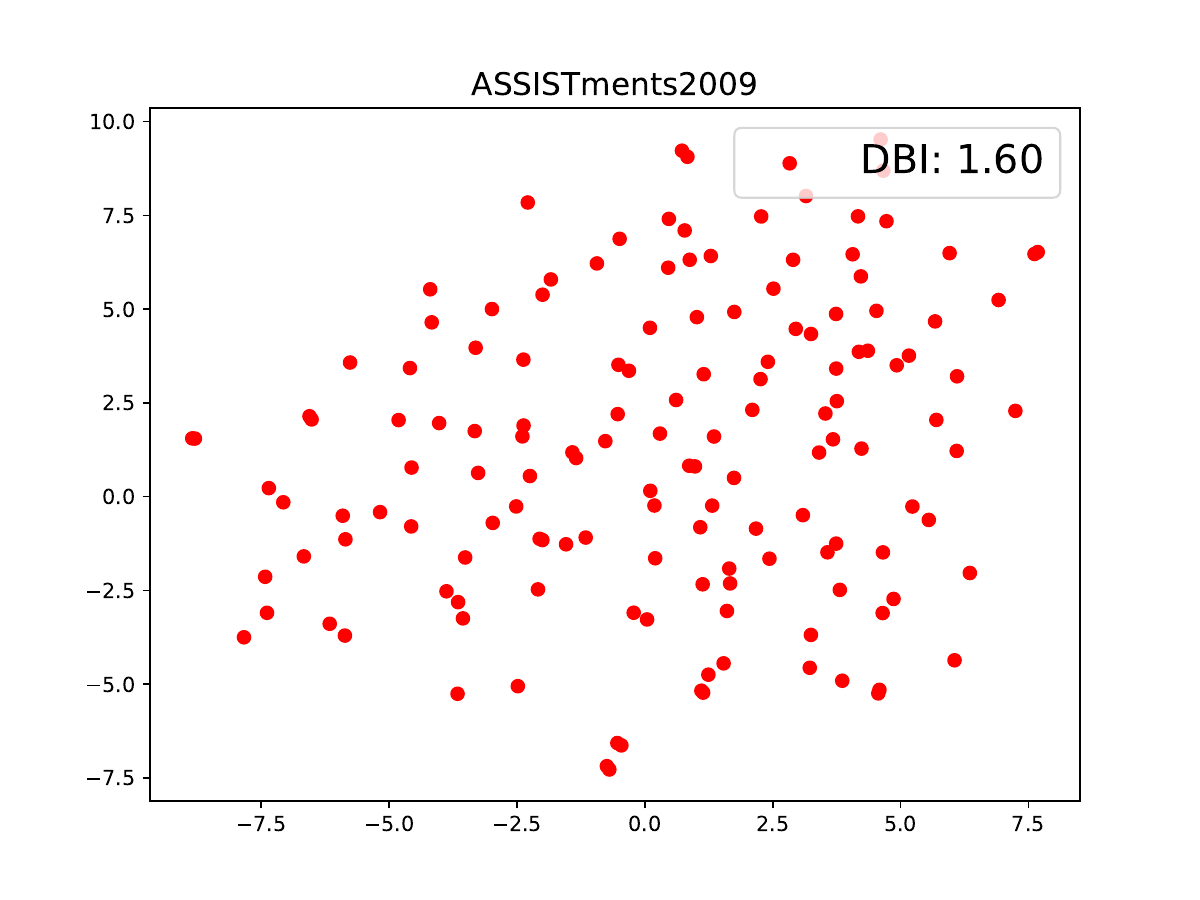}
           \vspace{-2mm}
            \caption{MoE embeddings}
    \end{subfigure}
    \begin{subfigure}[t]{0.325\textwidth}
           \centering
           \includegraphics[width=\columnwidth]{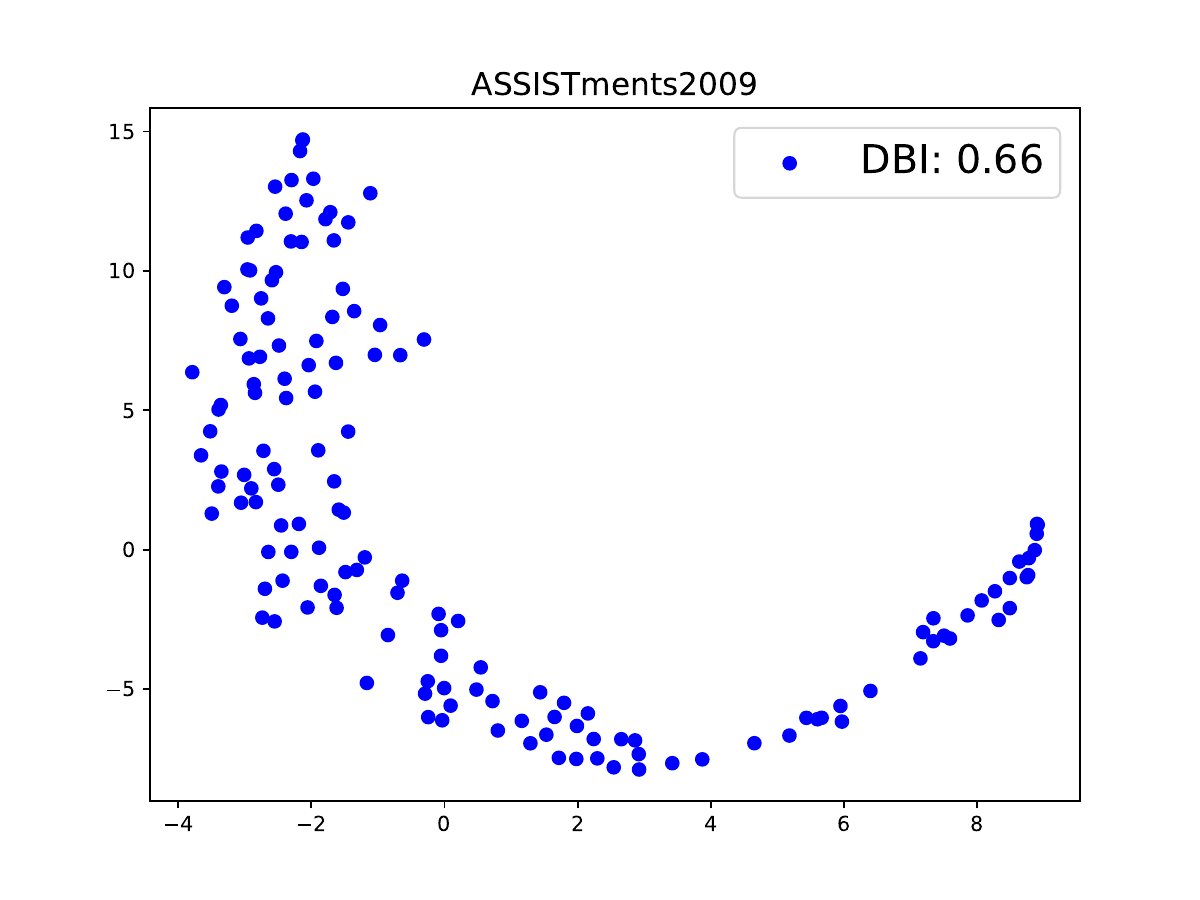}
           \vspace{-2mm}
            \caption{\our}
    \end{subfigure}
    \centering
    \vspace{-3mm}
    \caption{Comparing different distributions of embeddings in ASSIST09. The smaller the DBI value upper in figure, the better.}
    \vspace{-2mm}
    \label{fig:embedsDist09}
\end{figure*}

\subsubsection{Learner sequence modeling.}
In our framework, we mainly have two modules for learner sequence modeling: One is the knowledge tracing module to capture the learner's knowledge state from the learning history. The other is the sequence-based self-supervised pre-training (shortened as ``S3Pre-training'' since it's from the S3Rec method). We now explore the impact of them. The results are shown in Figure ~\ref{fig:AbOnSeq}. It was observed that removing knowledge tracing significantly deteriorates the model's performance. The reason is that each learner's learning journey is highly individualized, making concept recommendations heavily reliant on their current knowledge state. Therefore, the modeling of learners' knowledge states is crucial. Additionally, eliminating the sequence-based pre-training results in a notable decrease in performance. This highlights the importance of learning patterns within learner sequences, which is essential for the model to predict the next concept accurately.

\subsection{Consistency with Knowledge Structure}
\begin{figure}[htbp] 
    \centering
    \begin{subfigure}[t]{0.48\columnwidth}
           \centering
           \includegraphics[width=\columnwidth]{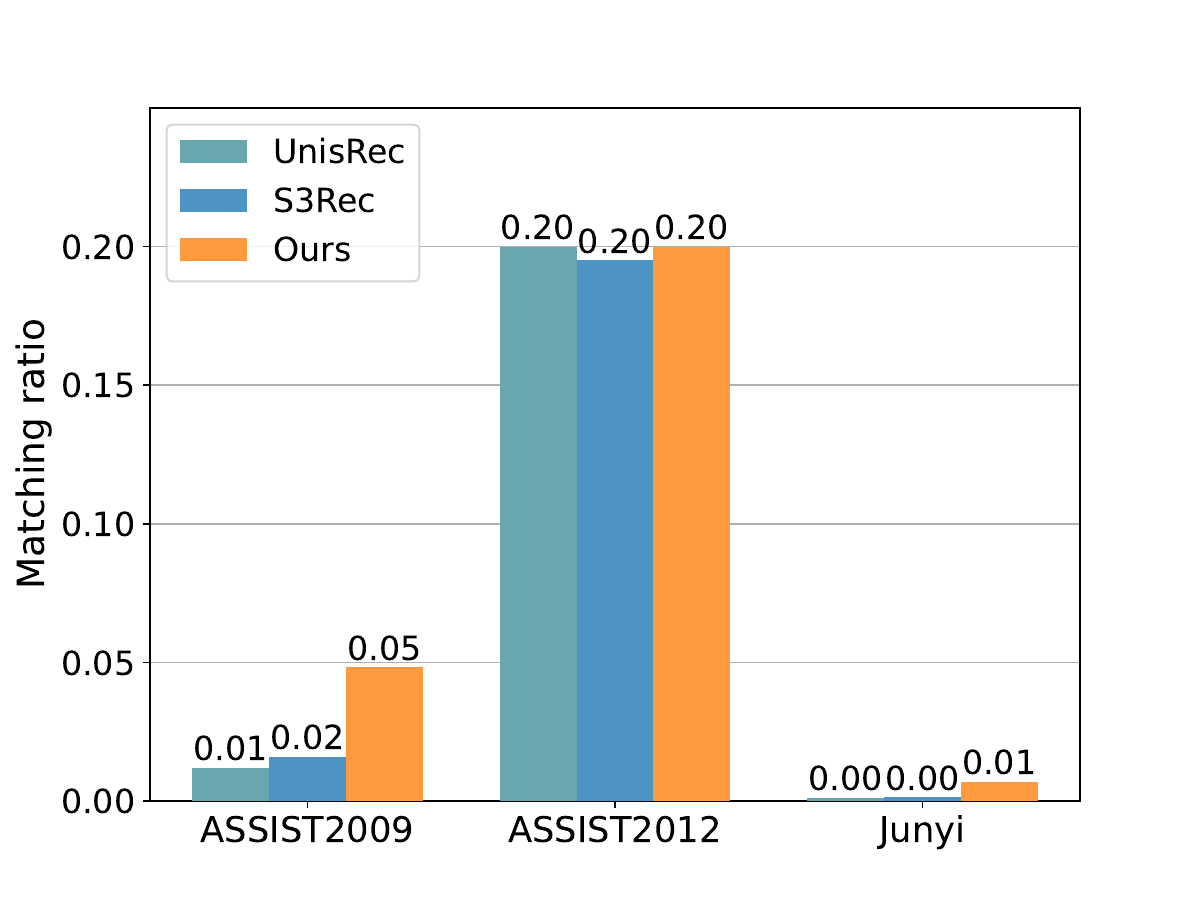}
            \caption{w/o learning style}
    \end{subfigure}
    \begin{subfigure}[t]{0.48\columnwidth}
           \centering
           \includegraphics[width=\columnwidth]{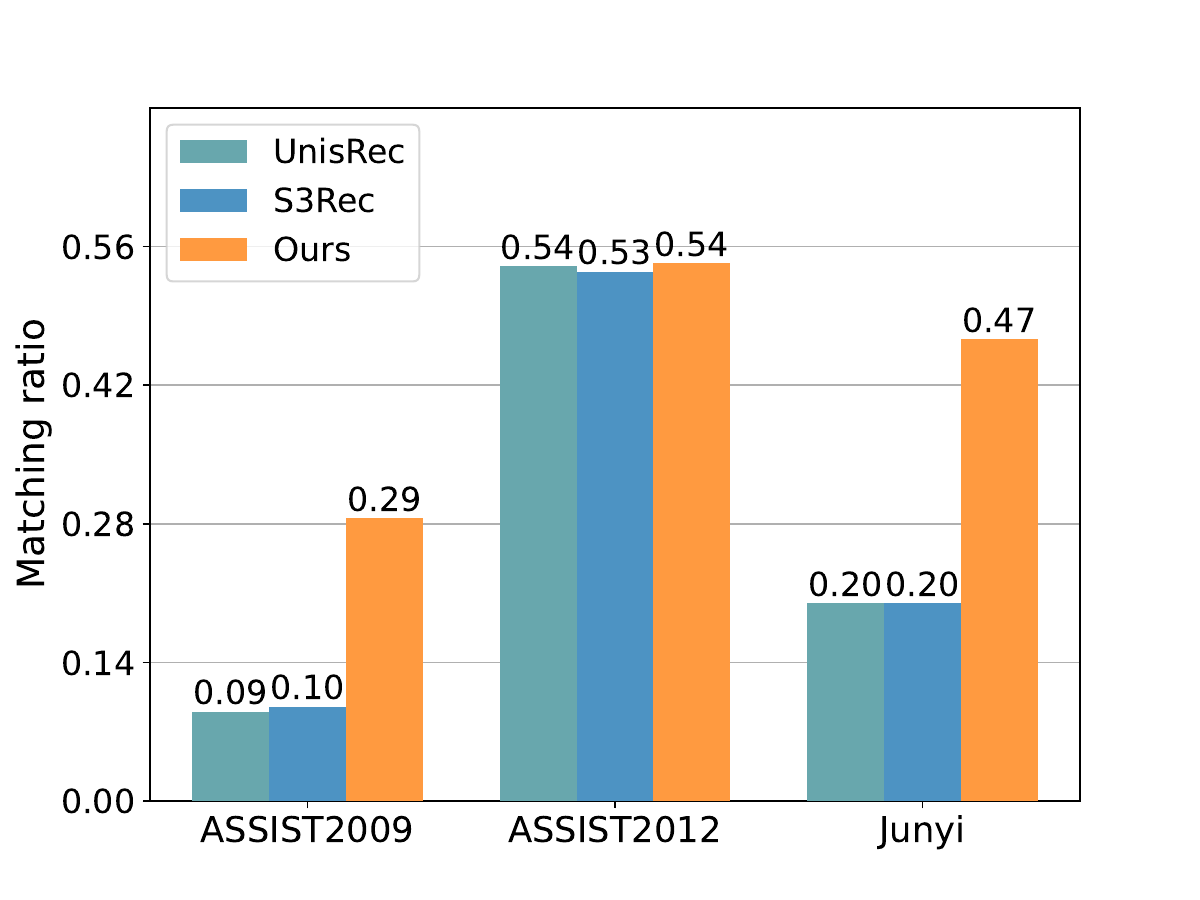}
            \caption{With learning style}
    \end{subfigure}
    \centering
    \vspace{-3mm}
    \caption{The matching ratios of the recommendation and graph structure.}
    \vspace{-2mm}
    \label{fig:agentModelStudy}
    \vspace{-2mm}
\end{figure}

The primary distinction between concept recommendation and product recommendation lies in that educational recommendation of concepts should adhere to the structure of human knowledge~\cite{liu2019exploiting}. This is manifested in data as the prerequisite relationships between concepts. In this section, we conducted experiments to evaluate if the models' recommendations are consistent with the knowledge graph. We compare the proportion of recommended concepts, $k_{t+1}$, having a prerequisite or successor relationship with the previous concept, $k_{t}$, on the knowledge graph. The results are displayed in Figure ~\ref{fig:agentModelStudy}. 

In sub-figure (a), we measure the share of recommendations that match the knowledge graph's structure on all learners. A recommendation for a learner is deemed consistent if it is a predecessor or successor to the previous concept. We express this as a ratio: $consistent ratio = \frac{\#consistent\ sample}{\#total\ sample}$  In sub-figure (b), instead of evaluating the consistency on all learners, we evaluate on the learners whose learning style is adhered to the graph. Specifically, we evaluate whether a learner's learning sequence aligns with the knowledge graph before making recommendations. We take each concept a learner has studied with the next as a pair and consider a pair ``consistent pair'' if they follow the graph's order. The ``consistency score'' is the ratio of consistent pairs to total pairs: $score = \frac{\#consistent\ pairs}{\#all\ pairs}$. A score over 0.5 suggests the learner's style aligns with the graph. We focus on the rate of consistent recommendations for this subset of learners.

The experiments indicate that our approach more frequently suggests concepts consistent with the graph's structure for all learners. This advantage is even more pronounced for those learners whose styles adhere to the graph's structure. This improvement stems from the structure-aware concept interpretation and the graph-based adapter, which enriches each concept's representation with its prerequisite relationships to other concepts.

\vspace{-3mm}
\subsection{Embedding Space Transformation}
To better understand the graph-based adapter's ability to adapt non-smooth, anisotropic text embedding spaces, we employed t-SNE to perform visualization of the vanilla text embeddings, graph-adapted embeddings, and MoE-adapted embeddings. The visualization comparison on ASSIST09 is shown in Figure ~\ref{fig:embedsDist09}. The other visualizations of ASSIST12 and Junyi are provided in Appendix ~\ref{sec:embd}.

\begin{table}[htbp]
\small
\centering

\caption{The DBI value of different representation learned in three datasets.}
\vspace{-1mm}
\label{tab:dbitable}
\scalebox{1.0}{
\setlength{\tabcolsep}{1mm}{
\begin{tabular}{lccc}
\toprule
\multicolumn{1}{c}{\textbf{Embeddings}} & ASSIST09& ASSIST12& Junyi \\ 
\midrule
Vanilla Text Embeddings  & 2.0257 & 2.3055   & 3.0333\\
MoE Adapted  & 1.6007  &  1.9199  & 2.0392\\
\cmidrule(r){1-4}
\our  & \textbf{0.6611}  & \textbf{0.3255}   & \textbf{1.7208}\\
\bottomrule
\end{tabular}
}}
\end{table}


From the illustrations provided, it is evident that the vanilla text embeddings and those adapted through MoE lack clear distribution patterns or clustering traits. However, embeddings that have undergone graph-based adaptation display pronounced manifold and clustering properties, making them more conducive to subsequent recommendation tasks.

To further illustrate the transformation effect, we calculate the Davies-Bouldin Index (DBI)~\cite{davies1979cluster}. It assesses the data separation state between clusters. A lower DBI indicates better-separated classes. We use K-Means to cluster the embeddings and calculate DBI as shown in Table ~\ref{tab:dbitable}. Our graph-based adapted embeddings achieve the lowest DBI scores, demonstrating a better clustering feature.

\section{Conclusion}
We present a structure and knowledge-aware representation framework for concept recommendation. Our approach involves generating textual descriptions for knowledge concepts using external information from LLMs and structural data from knowledge graphs. To address the challenge of converting non-smooth, anisotropic text encodings for concept recommendation, we have developed a graph-based adapter trained through contrastive learning. This marks the first effort to incorporate external knowledge from LLMs into the field of concept recommendation. Through thorough testing, our framework has proven effective in refining the textual information of concepts to enhance recommendation outcomes.


\bibliographystyle{ACM-Reference-Format}
\bibliography{main}


\begin{thebibliography}{33}


\ifx \showCODEN    \undefined \def \showCODEN     #1{\unskip}     \fi
\ifx \showDOI      \undefined \def \showDOI       #1{#1}\fi
\ifx \showISBNx    \undefined \def \showISBNx     #1{\unskip}     \fi
\ifx \showISBNxiii \undefined \def \showISBNxiii  #1{\unskip}     \fi
\ifx \showISSN     \undefined \def \showISSN      #1{\unskip}     \fi
\ifx \showLCCN     \undefined \def \showLCCN      #1{\unskip}     \fi
\ifx \shownote     \undefined \def \shownote      #1{#1}          \fi
\ifx \showarticletitle \undefined \def \showarticletitle #1{#1}   \fi
\ifx \showURL      \undefined \def \showURL       {\relax}        \fi
\providecommand\bibfield[2]{#2}
\providecommand\bibinfo[2]{#2}
\providecommand\natexlab[1]{#1}
\providecommand\showeprint[2][]{arXiv:#2}

\bibitem[Achiam et~al\mbox{.}(2023)]%
        {achiam2023gpt}
\bibfield{author}{\bibinfo{person}{Josh Achiam}, \bibinfo{person}{Steven Adler}, \bibinfo{person}{Sandhini Agarwal}, \bibinfo{person}{Lama Ahmad}, \bibinfo{person}{Ilge Akkaya}, \bibinfo{person}{Florencia~Leoni Aleman}, \bibinfo{person}{Diogo Almeida}, \bibinfo{person}{Janko Altenschmidt}, \bibinfo{person}{Sam Altman}, \bibinfo{person}{Shyamal Anadkat}, {et~al\mbox{.}}} \bibinfo{year}{2023}\natexlab{}.
\newblock \showarticletitle{Gpt-4 technical report}.
\newblock \bibinfo{journal}{\emph{arXiv preprint arXiv:2303.08774}} (\bibinfo{year}{2023}).
\newblock


\bibitem[Beltagy et~al\mbox{.}(2020)]%
        {beltagy2020longformer}
\bibfield{author}{\bibinfo{person}{Iz Beltagy}, \bibinfo{person}{Matthew~E Peters}, {and} \bibinfo{person}{Arman Cohan}.} \bibinfo{year}{2020}\natexlab{}.
\newblock \showarticletitle{Longformer: The long-document transformer}.
\newblock \bibinfo{journal}{\emph{arXiv preprint arXiv:2004.05150}} (\bibinfo{year}{2020}).
\newblock


\bibitem[Chung et~al\mbox{.}(2014)]%
        {chung2014empirical}
\bibfield{author}{\bibinfo{person}{Junyoung Chung}, \bibinfo{person}{Caglar Gulcehre}, \bibinfo{person}{KyungHyun Cho}, {and} \bibinfo{person}{Yoshua Bengio}.} \bibinfo{year}{2014}\natexlab{}.
\newblock \showarticletitle{Empirical evaluation of gated recurrent neural networks on sequence modeling}.
\newblock \bibinfo{journal}{\emph{arXiv preprint arXiv:1412.3555}} (\bibinfo{year}{2014}).
\newblock


\bibitem[Davies and Bouldin(1979)]%
        {davies1979cluster}
\bibfield{author}{\bibinfo{person}{David~L Davies} {and} \bibinfo{person}{Donald~W Bouldin}.} \bibinfo{year}{1979}\natexlab{}.
\newblock \showarticletitle{A cluster separation measure}.
\newblock \bibinfo{journal}{\emph{IEEE transactions on pattern analysis and machine intelligence}} \bibinfo{number}{2} (\bibinfo{year}{1979}), \bibinfo{pages}{224--227}.
\newblock


\bibitem[Devlin et~al\mbox{.}(2018)]%
        {devlin2018bert}
\bibfield{author}{\bibinfo{person}{Jacob Devlin}, \bibinfo{person}{Ming-Wei Chang}, \bibinfo{person}{Kenton Lee}, {and} \bibinfo{person}{Kristina Toutanova}.} \bibinfo{year}{2018}\natexlab{}.
\newblock \showarticletitle{Bert: Pre-training of deep bidirectional transformers for language understanding}.
\newblock \bibinfo{journal}{\emph{arXiv preprint arXiv:1810.04805}} (\bibinfo{year}{2018}).
\newblock


\bibitem[Ding et~al\mbox{.}(2021)]%
        {ding2021zero}
\bibfield{author}{\bibinfo{person}{Hao Ding}, \bibinfo{person}{Yifei Ma}, \bibinfo{person}{Anoop Deoras}, \bibinfo{person}{Yuyang Wang}, {and} \bibinfo{person}{Hao Wang}.} \bibinfo{year}{2021}\natexlab{}.
\newblock \showarticletitle{Zero-shot recommender systems}.
\newblock \bibinfo{journal}{\emph{arXiv preprint arXiv:2105.08318}} (\bibinfo{year}{2021}).
\newblock


\bibitem[Geng et~al\mbox{.}(2022)]%
        {geng2022recommendation}
\bibfield{author}{\bibinfo{person}{Shijie Geng}, \bibinfo{person}{Shuchang Liu}, \bibinfo{person}{Zuohui Fu}, \bibinfo{person}{Yingqiang Ge}, {and} \bibinfo{person}{Yongfeng Zhang}.} \bibinfo{year}{2022}\natexlab{}.
\newblock \showarticletitle{Recommendation as language processing (rlp): A unified pretrain, personalized prompt \& predict paradigm (p5)}. In \bibinfo{booktitle}{\emph{Proceedings of the 16th ACM Conference on Recommender Systems}}. \bibinfo{pages}{299--315}.
\newblock


\bibitem[Gong et~al\mbox{.}(2020)]%
        {gong2020attentional}
\bibfield{author}{\bibinfo{person}{Jibing Gong}, \bibinfo{person}{Shen Wang}, \bibinfo{person}{Jinlong Wang}, \bibinfo{person}{Wenzheng Feng}, \bibinfo{person}{Hao Peng}, \bibinfo{person}{Jie Tang}, {and} \bibinfo{person}{Philip~S Yu}.} \bibinfo{year}{2020}\natexlab{}.
\newblock \showarticletitle{Attentional graph convolutional networks for knowledge concept recommendation in moocs in a heterogeneous view}. In \bibinfo{booktitle}{\emph{Proceedings of the 43rd international ACM SIGIR conference on research and development in information retrieval}}. \bibinfo{pages}{79--88}.
\newblock


\bibitem[Hou et~al\mbox{.}(2022)]%
        {hou2022towards}
\bibfield{author}{\bibinfo{person}{Yupeng Hou}, \bibinfo{person}{Shanlei Mu}, \bibinfo{person}{Wayne~Xin Zhao}, \bibinfo{person}{Yaliang Li}, \bibinfo{person}{Bolin Ding}, {and} \bibinfo{person}{Ji-Rong Wen}.} \bibinfo{year}{2022}\natexlab{}.
\newblock \showarticletitle{Towards universal sequence representation learning for recommender systems}. In \bibinfo{booktitle}{\emph{Proceedings of the 28th ACM SIGKDD Conference on Knowledge Discovery and Data Mining}}. \bibinfo{pages}{585--593}.
\newblock


\bibitem[Kang and McAuley(2018)]%
        {kang2018self}
\bibfield{author}{\bibinfo{person}{Wang-Cheng Kang} {and} \bibinfo{person}{Julian McAuley}.} \bibinfo{year}{2018}\natexlab{}.
\newblock \showarticletitle{Self-attentive sequential recommendation}. In \bibinfo{booktitle}{\emph{2018 IEEE international conference on data mining (ICDM)}}. IEEE, \bibinfo{pages}{197--206}.
\newblock


\bibitem[Kipf and Welling(2016)]%
        {kipf2016semi}
\bibfield{author}{\bibinfo{person}{Thomas~N Kipf} {and} \bibinfo{person}{Max Welling}.} \bibinfo{year}{2016}\natexlab{}.
\newblock \showarticletitle{Semi-supervised classification with graph convolutional networks}.
\newblock \bibinfo{journal}{\emph{arXiv preprint arXiv:1609.02907}} (\bibinfo{year}{2016}).
\newblock


\bibitem[Lewis et~al\mbox{.}(2019)]%
        {lewis2019bart}
\bibfield{author}{\bibinfo{person}{Mike Lewis}, \bibinfo{person}{Yinhan Liu}, \bibinfo{person}{Naman Goyal}, \bibinfo{person}{Marjan Ghazvininejad}, \bibinfo{person}{Abdelrahman Mohamed}, \bibinfo{person}{Omer Levy}, \bibinfo{person}{Ves Stoyanov}, {and} \bibinfo{person}{Luke Zettlemoyer}.} \bibinfo{year}{2019}\natexlab{}.
\newblock \showarticletitle{Bart: Denoising sequence-to-sequence pre-training for natural language generation, translation, and comprehension}.
\newblock \bibinfo{journal}{\emph{arXiv preprint arXiv:1910.13461}} (\bibinfo{year}{2019}).
\newblock


\bibitem[Li et~al\mbox{.}(2020)]%
        {li2020sentence}
\bibfield{author}{\bibinfo{person}{Bohan Li}, \bibinfo{person}{Hao Zhou}, \bibinfo{person}{Junxian He}, \bibinfo{person}{Mingxuan Wang}, \bibinfo{person}{Yiming Yang}, {and} \bibinfo{person}{Lei Li}.} \bibinfo{year}{2020}\natexlab{}.
\newblock \showarticletitle{On the sentence embeddings from pre-trained language models}.
\newblock \bibinfo{journal}{\emph{arXiv preprint arXiv:2011.05864}} (\bibinfo{year}{2020}).
\newblock


\bibitem[Li et~al\mbox{.}(2023a)]%
        {li2023text}
\bibfield{author}{\bibinfo{person}{Jiacheng Li}, \bibinfo{person}{Ming Wang}, \bibinfo{person}{Jin Li}, \bibinfo{person}{Jinmiao Fu}, \bibinfo{person}{Xin Shen}, \bibinfo{person}{Jingbo Shang}, {and} \bibinfo{person}{Julian McAuley}.} \bibinfo{year}{2023}\natexlab{a}.
\newblock \showarticletitle{Text Is All You Need: Learning Language Representations for Sequential Recommendation}.
\newblock \bibinfo{journal}{\emph{arXiv preprint arXiv:2305.13731}} (\bibinfo{year}{2023}).
\newblock


\bibitem[Li et~al\mbox{.}(2018)]%
        {li2018deeper}
\bibfield{author}{\bibinfo{person}{Qimai Li}, \bibinfo{person}{Zhichao Han}, {and} \bibinfo{person}{Xiao-Ming Wu}.} \bibinfo{year}{2018}\natexlab{}.
\newblock \showarticletitle{Deeper insights into graph convolutional networks for semi-supervised learning}. In \bibinfo{booktitle}{\emph{Proceedings of the AAAI conference on artificial intelligence}}, Vol.~\bibinfo{volume}{32}.
\newblock


\bibitem[Li et~al\mbox{.}(2023b)]%
        {li2023graph}
\bibfield{author}{\bibinfo{person}{Qingyao Li}, \bibinfo{person}{Wei Xia}, \bibinfo{person}{Li'ang Yin}, \bibinfo{person}{Jian Shen}, \bibinfo{person}{Renting Rui}, \bibinfo{person}{Weinan Zhang}, \bibinfo{person}{Xianyu Chen}, \bibinfo{person}{Ruiming Tang}, {and} \bibinfo{person}{Yong Yu}.} \bibinfo{year}{2023}\natexlab{b}.
\newblock \showarticletitle{Graph Enhanced Hierarchical Reinforcement Learning for Goal-oriented Learning Path Recommendation}. In \bibinfo{booktitle}{\emph{Proceedings of the 32nd ACM International Conference on Information and Knowledge Management}}. \bibinfo{pages}{1318--1327}.
\newblock


\bibitem[Lin et~al\mbox{.}(2023)]%
        {lin2023can}
\bibfield{author}{\bibinfo{person}{Jianghao Lin}, \bibinfo{person}{Xinyi Dai}, \bibinfo{person}{Yunjia Xi}, \bibinfo{person}{Weiwen Liu}, \bibinfo{person}{Bo Chen}, \bibinfo{person}{Xiangyang Li}, \bibinfo{person}{Chenxu Zhu}, \bibinfo{person}{Huifeng Guo}, \bibinfo{person}{Yong Yu}, \bibinfo{person}{Ruiming Tang}, {et~al\mbox{.}}} \bibinfo{year}{2023}\natexlab{}.
\newblock \showarticletitle{How Can Recommender Systems Benefit from Large Language Models: A Survey}.
\newblock \bibinfo{journal}{\emph{arXiv preprint arXiv:2306.05817}} (\bibinfo{year}{2023}).
\newblock


\bibitem[Lin et~al\mbox{.}(2021)]%
        {lin2021adaptive}
\bibfield{author}{\bibinfo{person}{Yuanguo Lin}, \bibinfo{person}{Shibo Feng}, \bibinfo{person}{Fan Lin}, \bibinfo{person}{Wenhua Zeng}, \bibinfo{person}{Yong Liu}, {and} \bibinfo{person}{Pengcheng Wu}.} \bibinfo{year}{2021}\natexlab{}.
\newblock \showarticletitle{Adaptive course recommendation in MOOCs}.
\newblock \bibinfo{journal}{\emph{Knowledge-Based Systems}}  \bibinfo{volume}{224} (\bibinfo{year}{2021}), \bibinfo{pages}{107085}.
\newblock


\bibitem[Liu et~al\mbox{.}(2018)]%
        {liu2018finding}
\bibfield{author}{\bibinfo{person}{Qi Liu}, \bibinfo{person}{Zai Huang}, \bibinfo{person}{Zhenya Huang}, \bibinfo{person}{Chuanren Liu}, \bibinfo{person}{Enhong Chen}, \bibinfo{person}{Yu Su}, {and} \bibinfo{person}{Guoping Hu}.} \bibinfo{year}{2018}\natexlab{}.
\newblock \showarticletitle{Finding similar exercises in online education systems}. In \bibinfo{booktitle}{\emph{Proceedings of the 24th ACM SIGKDD International Conference on Knowledge Discovery \& Data Mining}}. \bibinfo{pages}{1821--1830}.
\newblock


\bibitem[Liu et~al\mbox{.}(2019)]%
        {liu2019exploiting}
\bibfield{author}{\bibinfo{person}{Qi Liu}, \bibinfo{person}{Shiwei Tong}, \bibinfo{person}{Chuanren Liu}, \bibinfo{person}{Hongke Zhao}, \bibinfo{person}{Enhong Chen}, \bibinfo{person}{Haiping Ma}, {and} \bibinfo{person}{Shijin Wang}.} \bibinfo{year}{2019}\natexlab{}.
\newblock \showarticletitle{Exploiting cognitive structure for adaptive learning}. In \bibinfo{booktitle}{\emph{Proceedings of the 25th ACM SIGKDD International Conference on Knowledge Discovery \& Data Mining}}. \bibinfo{pages}{627--635}.
\newblock


\bibitem[Liu et~al\mbox{.}(2022)]%
        {liu2022pykt}
\bibfield{author}{\bibinfo{person}{Zitao Liu}, \bibinfo{person}{Qiongqiong Liu}, \bibinfo{person}{Jiahao Chen}, \bibinfo{person}{Shuyan Huang}, \bibinfo{person}{Jiliang Tang}, {and} \bibinfo{person}{Weiqi Luo}.} \bibinfo{year}{2022}\natexlab{}.
\newblock \showarticletitle{pyKT: a python library to benchmark deep learning based knowledge tracing models}.
\newblock \bibinfo{journal}{\emph{Advances in Neural Information Processing Systems}}  \bibinfo{volume}{35} (\bibinfo{year}{2022}), \bibinfo{pages}{18542--18555}.
\newblock


\bibitem[Long et~al\mbox{.}(2022)]%
        {long2022improving}
\bibfield{author}{\bibinfo{person}{Ting Long}, \bibinfo{person}{Jiarui Qin}, \bibinfo{person}{Jian Shen}, \bibinfo{person}{Weinan Zhang}, \bibinfo{person}{Wei Xia}, \bibinfo{person}{Ruiming Tang}, \bibinfo{person}{Xiuqiang He}, {and} \bibinfo{person}{Yong Yu}.} \bibinfo{year}{2022}\natexlab{}.
\newblock \showarticletitle{Improving knowledge tracing with collaborative information}. In \bibinfo{booktitle}{\emph{Proceedings of the fifteenth ACM international conference on web search and data mining}}. \bibinfo{pages}{599--607}.
\newblock


\bibitem[Nakagawa et~al\mbox{.}(2019)]%
        {nakagawa2019graph}
\bibfield{author}{\bibinfo{person}{Hiromi Nakagawa}, \bibinfo{person}{Yusuke Iwasawa}, {and} \bibinfo{person}{Yutaka Matsuo}.} \bibinfo{year}{2019}\natexlab{}.
\newblock \showarticletitle{Graph-based knowledge tracing: modeling student proficiency using graph neural network}. In \bibinfo{booktitle}{\emph{IEEE/WIC/ACM International Conference on Web Intelligence}}. \bibinfo{pages}{156--163}.
\newblock


\bibitem[Oord et~al\mbox{.}(2018)]%
        {oord2018representation}
\bibfield{author}{\bibinfo{person}{Aaron van~den Oord}, \bibinfo{person}{Yazhe Li}, {and} \bibinfo{person}{Oriol Vinyals}.} \bibinfo{year}{2018}\natexlab{}.
\newblock \showarticletitle{Representation learning with contrastive predictive coding}.
\newblock \bibinfo{journal}{\emph{arXiv preprint arXiv:1807.03748}} (\bibinfo{year}{2018}).
\newblock


\bibitem[Piech et~al\mbox{.}(2015)]%
        {piech2015deep}
\bibfield{author}{\bibinfo{person}{Chris Piech}, \bibinfo{person}{Jonathan Bassen}, \bibinfo{person}{Jonathan Huang}, \bibinfo{person}{Surya Ganguli}, \bibinfo{person}{Mehran Sahami}, \bibinfo{person}{Leonidas~J Guibas}, {and} \bibinfo{person}{Jascha Sohl-Dickstein}.} \bibinfo{year}{2015}\natexlab{}.
\newblock \showarticletitle{Deep knowledge tracing}.
\newblock \bibinfo{journal}{\emph{Advances in neural information processing systems}}  \bibinfo{volume}{28} (\bibinfo{year}{2015}).
\newblock


\bibitem[Sun et~al\mbox{.}(2019)]%
        {sun2019bert4rec}
\bibfield{author}{\bibinfo{person}{Fei Sun}, \bibinfo{person}{Jun Liu}, \bibinfo{person}{Jian Wu}, \bibinfo{person}{Changhua Pei}, \bibinfo{person}{Xiao Lin}, \bibinfo{person}{Wenwu Ou}, {and} \bibinfo{person}{Peng Jiang}.} \bibinfo{year}{2019}\natexlab{}.
\newblock \showarticletitle{BERT4Rec: Sequential recommendation with bidirectional encoder representations from transformer}. In \bibinfo{booktitle}{\emph{Proceedings of the 28th ACM international conference on information and knowledge management}}. \bibinfo{pages}{1441--1450}.
\newblock


\bibitem[Vaswani et~al\mbox{.}(2017)]%
        {vaswani2017attention}
\bibfield{author}{\bibinfo{person}{Ashish Vaswani}, \bibinfo{person}{Noam Shazeer}, \bibinfo{person}{Niki Parmar}, \bibinfo{person}{Jakob Uszkoreit}, \bibinfo{person}{Llion Jones}, \bibinfo{person}{Aidan~N Gomez}, \bibinfo{person}{{\L}ukasz Kaiser}, {and} \bibinfo{person}{Illia Polosukhin}.} \bibinfo{year}{2017}\natexlab{}.
\newblock \showarticletitle{Attention is all you need}.
\newblock \bibinfo{journal}{\emph{Advances in neural information processing systems}}  \bibinfo{volume}{30} (\bibinfo{year}{2017}).
\newblock


\bibitem[Wu et~al\mbox{.}(2021)]%
        {wu2021self}
\bibfield{author}{\bibinfo{person}{Jiancan Wu}, \bibinfo{person}{Xiang Wang}, \bibinfo{person}{Fuli Feng}, \bibinfo{person}{Xiangnan He}, \bibinfo{person}{Liang Chen}, \bibinfo{person}{Jianxun Lian}, {and} \bibinfo{person}{Xing Xie}.} \bibinfo{year}{2021}\natexlab{}.
\newblock \showarticletitle{Self-supervised graph learning for recommendation}. In \bibinfo{booktitle}{\emph{Proceedings of the 44th international ACM SIGIR conference on research and development in information retrieval}}. \bibinfo{pages}{726--735}.
\newblock


\bibitem[Yang et~al\mbox{.}(2021)]%
        {yang2021gikt}
\bibfield{author}{\bibinfo{person}{Yang Yang}, \bibinfo{person}{Jian Shen}, \bibinfo{person}{Yanru Qu}, \bibinfo{person}{Yunfei Liu}, \bibinfo{person}{Kerong Wang}, \bibinfo{person}{Yaoming Zhu}, \bibinfo{person}{Weinan Zhang}, {and} \bibinfo{person}{Yong Yu}.} \bibinfo{year}{2021}\natexlab{}.
\newblock \showarticletitle{GIKT: a graph-based interaction model for knowledge tracing}. In \bibinfo{booktitle}{\emph{Machine Learning and Knowledge Discovery in Databases: European Conference, ECML PKDD 2020, Ghent, Belgium, September 14--18, 2020, Proceedings, Part I}}. Springer, \bibinfo{pages}{299--315}.
\newblock


\bibitem[Yu et~al\mbox{.}(2020)]%
        {yu2020mooccube}
\bibfield{author}{\bibinfo{person}{Jifan Yu}, \bibinfo{person}{Gan Luo}, \bibinfo{person}{Tong Xiao}, \bibinfo{person}{Qingyang Zhong}, \bibinfo{person}{Yuquan Wang}, \bibinfo{person}{Wenzheng Feng}, \bibinfo{person}{Junyi Luo}, \bibinfo{person}{Chenyu Wang}, \bibinfo{person}{Lei Hou}, \bibinfo{person}{Juanzi Li}, {et~al\mbox{.}}} \bibinfo{year}{2020}\natexlab{}.
\newblock \showarticletitle{MOOCCube: a large-scale data repository for NLP applications in MOOCs}. In \bibinfo{booktitle}{\emph{Proceedings of the 58th annual meeting of the association for computational linguistics}}. \bibinfo{pages}{3135--3142}.
\newblock


\bibitem[Yu et~al\mbox{.}(2023)]%
        {yu2023graph}
\bibfield{author}{\bibinfo{person}{Mei Yu}, \bibinfo{person}{Zhaoyuan Ding}, \bibinfo{person}{Jian Yu}, \bibinfo{person}{Wenbin Zhang}, \bibinfo{person}{Ming Yang}, {and} \bibinfo{person}{Mankun Zhao}.} \bibinfo{year}{2023}\natexlab{}.
\newblock \showarticletitle{Graph Contrastive Learning with Adaptive Augmentation for Knowledge Concept Recommendation}. In \bibinfo{booktitle}{\emph{2023 26th International Conference on Computer Supported Cooperative Work in Design (CSCWD)}}. IEEE, \bibinfo{pages}{1281--1286}.
\newblock


\bibitem[Zhang et~al\mbox{.}(2019)]%
        {zhang2019feature}
\bibfield{author}{\bibinfo{person}{Tingting Zhang}, \bibinfo{person}{Pengpeng Zhao}, \bibinfo{person}{Yanchi Liu}, \bibinfo{person}{Victor~S Sheng}, \bibinfo{person}{Jiajie Xu}, \bibinfo{person}{Deqing Wang}, \bibinfo{person}{Guanfeng Liu}, \bibinfo{person}{Xiaofang Zhou}, {et~al\mbox{.}}} \bibinfo{year}{2019}\natexlab{}.
\newblock \showarticletitle{Feature-level Deeper Self-Attention Network for Sequential Recommendation.}. In \bibinfo{booktitle}{\emph{IJCAI}}. \bibinfo{pages}{4320--4326}.
\newblock


\bibitem[Zhou et~al\mbox{.}(2020)]%
        {zhou2020s3}
\bibfield{author}{\bibinfo{person}{Kun Zhou}, \bibinfo{person}{Hui Wang}, \bibinfo{person}{Wayne~Xin Zhao}, \bibinfo{person}{Yutao Zhu}, \bibinfo{person}{Sirui Wang}, \bibinfo{person}{Fuzheng Zhang}, \bibinfo{person}{Zhongyuan Wang}, {and} \bibinfo{person}{Ji-Rong Wen}.} \bibinfo{year}{2020}\natexlab{}.
\newblock \showarticletitle{S3-rec: Self-supervised learning for sequential recommendation with mutual information maximization}. In \bibinfo{booktitle}{\emph{Proceedings of the 29th ACM international conference on information \& knowledge management}}. \bibinfo{pages}{1893--1902}.
\newblock


\end{thebibliography}

\appendix

\section{Algorithm} \label{sec:algo}

We present the detailed pre-training and fine-tuning procedure in Algorithm ~\ref{algo}.











\begin{algorithm}[htbp]

\caption{Pre-training and Fine-tuning Framework}

\label{algo}
\textbf{Input}: Knowledge graph $G$, learner sequences $D_{\mathrm{train}}$, Random initialized graph-based adapter $F_{g}$, knowledge tracing network $F_{kt}$, Transformer blocks $F_{tran}$ \\
\textbf{Hyper-parameters}: $n_{\mathrm{epoch}}$, edge masking ratio $\gamma$ \\
\textbf{Output}: Trained $F_{g}$, $F_{kt}$, $F_{trans}$

\begin{algorithmic}[1]

\STATE $D_{\mathrm{train}} \leftarrow \mathrm{LLMTextEnhance}(D_{\mathrm{train}})$ \\
\STATE $D_{\mathrm{train}} \leftarrow \mathrm{LLMTextEncoding}(D_{\mathrm{train}})$ \\
\textbf{\emph{Pre-training}} \\

\FOR{$n$ in $n_{\mathrm{epoch}}$}
    \STATE $v_1(G), v_2(G) \leftarrow \mathrm{EdgeMask}(G, \gamma)$ \\
    \STATE $\mathcal{L}_{ssl}^{G} \leftarrow \mathrm{Contrastive Learning}(v_1(G), v_2(G), F_{g})$ \\
    \STATE $F_{g} \leftarrow \mathrm{Update}(\mathcal{L}_{ssl}^{G}, F_{g})$ \\
\ENDFOR 

\FOR{$n$ in $n_{\mathrm{epoch}}$}
    \STATE $\mathcal{L}^{KT} \leftarrow \mathrm{KnolwedgeTracingTask}(F_{kt}, D_{\mathrm{train}})$ \\
    \STATE $F_{kt} \leftarrow \mathrm{Update}(\mathcal{L}^{kt}, F_{kt})$ \\
\ENDFOR 

\FOR{$n$ in $n_{\mathrm{epoch}}$}
    \STATE $D^{MIP} \leftarrow \mathrm{MaskItem}(D_{\mathrm{train}})$ \\
    \STATE $D^{MSP} \leftarrow \mathrm{MaskSegment}(D_{\mathrm{train}})$ \\
    \STATE $D^{MAP} \leftarrow \mathrm{MaskAttribute}(D_{\mathrm{train}})$ \\
    \STATE $D^{AAP} \leftarrow \mathrm{MaskAssociatedAttribute}(D_{\mathrm{train}})$ \\

    \STATE $\mathcal{L}^{MIP} \leftarrow \mathrm{MaskItemPredict}(D^{MIP},F_{trans},F_{g},F_{kt})$ \\
    \STATE $\mathcal{L}^{MSP} \leftarrow \mathrm{MaskSegmentPredict}(D^{MSP}, F_{trans},F_{g},F_{kt})$ \\
    \STATE $\mathcal{L}^{MAP} \leftarrow \mathrm{MaskAttributePredict}(D^{MAP}, F_{trans},F_{g},F_{kt})$ \\
    \STATE $\mathcal{L}^{AAP} \leftarrow \mathrm{AttributePredict}(D^{AAP}, F_{trans},F_{g},F_{kt})$ \\
    \STATE $\mathcal{L}_{ssl}^{seq} \leftarrow \mathcal{L}^{MIP} + \mathcal{L}^{MSP}+ \mathcal{L}^{MAP}+\mathcal{L}^{AAP} $\\
    \STATE $F_{trans},F_{g},F_{kt}  \leftarrow \mathrm{Update}(\mathcal{L}_{ssl}^{seq}, F_{trans}, F_{g}, F_{kt})$ \\
\ENDFOR

\textbf{\emph{Fine-tuning} }\\
\FOR{$n$ in $n_{\mathrm{epoch}}$}
    \STATE $\mathcal{L}_{rec} \leftarrow \mathrm{NextConceptPred}(F_{trans},F_{g},F_{kt}, D_{\mathrm{train}})$ \\
    \STATE $F_{trans},F_{g},F_{kt} \leftarrow \mathrm{Update}(\mathcal{L}_{rec}, F_{trans},F_{g},F_{kt} )$ \\
\ENDFOR 

\STATE \textbf{return} $F_{trans},F_{g},F_{kt}$ \\

\end{algorithmic}
\end{algorithm}

\section{Additional Experiments Details}
\subsection{Implementation Details.}
For baselines purely based on heterogeneous graph-based concept recommendation, we constructed a bipartite graph of learner-concept with two types of edges: correct-answer edges and wrong-answer edges. In our method, \our, we implemented three layers of attention blocks. The sizes for the ID embedding, answer embedding, and graph-adapted embedding were all set to 64. The maximum sequence length was configured to be 200, and the batch size was set at 256. The masking ratio $\gamma$ is tuned in $\{0.1, 0.2, 0.3, 0.4, 0.5\}$. We utilize GPT-3.5 to do the concept interpretation. 

\subsection{Impact of edge masking ratio $\gamma$}
\begin{figure*}[htbp]
    \centering
    \begin{subfigure}[t]{0.325\textwidth}
           \centering
           \includegraphics[width=\columnwidth]{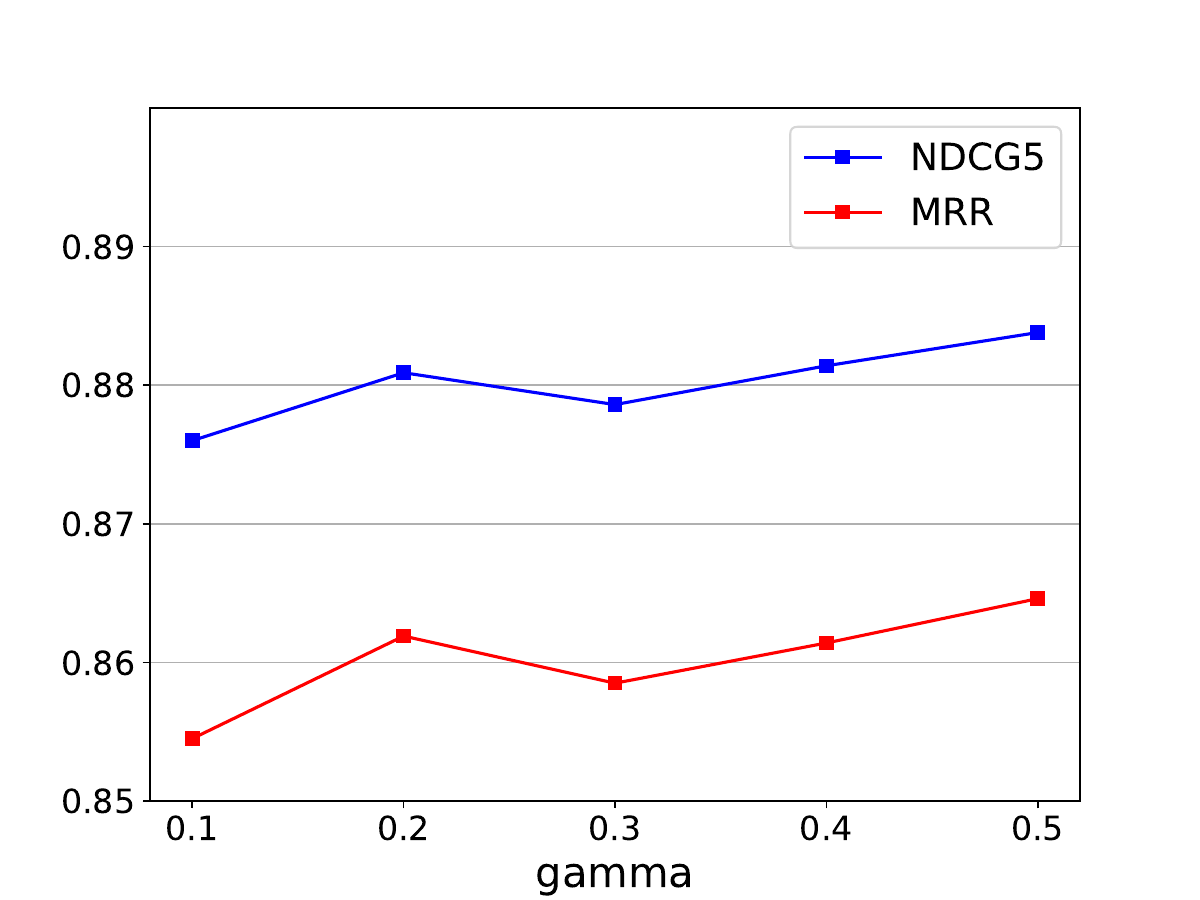}
            \caption{ASSIST09}
    \end{subfigure}
    \begin{subfigure}[t]{0.325\textwidth}
           \centering
           \includegraphics[width=\columnwidth]{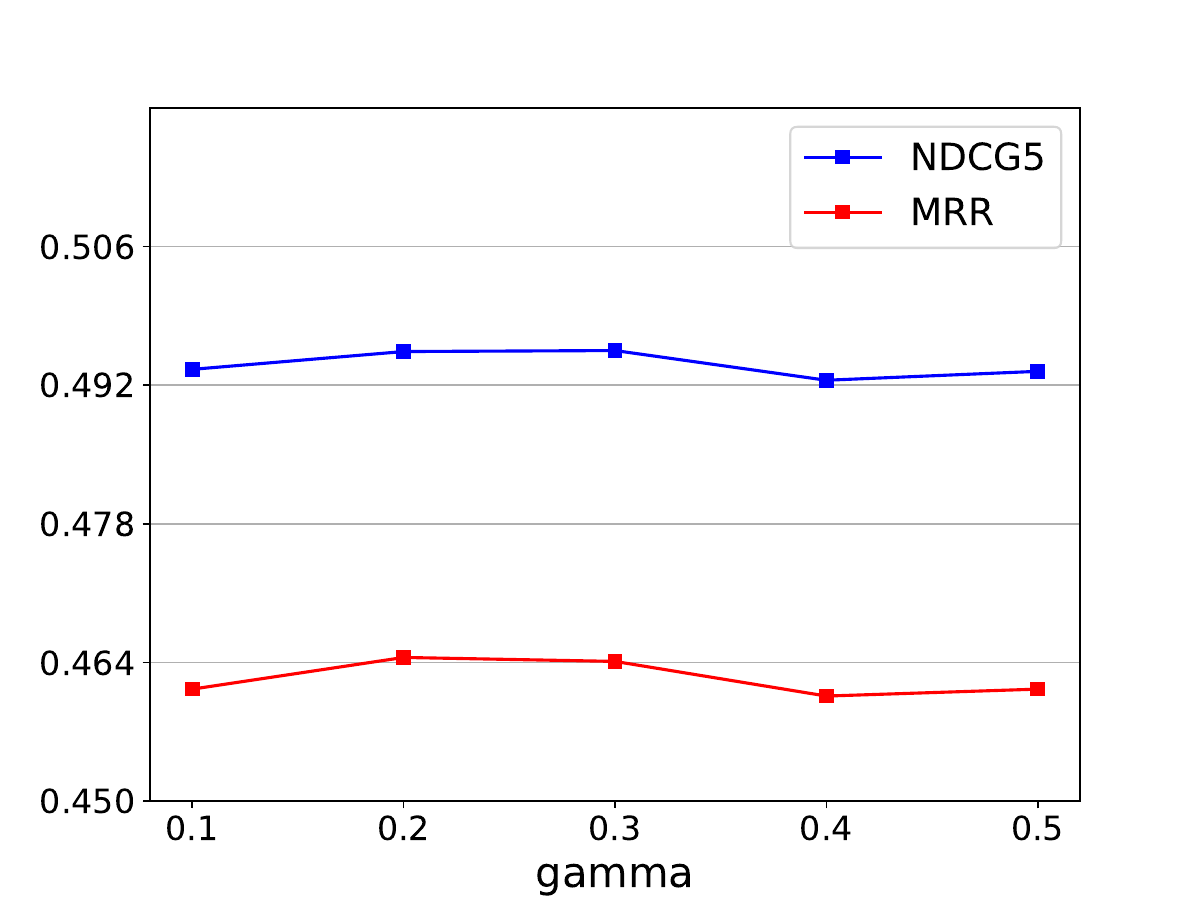}
            \caption{ASSIST12}
    \end{subfigure}
    \begin{subfigure}[t]{0.325\textwidth}
           \centering
           \includegraphics[width=\columnwidth]{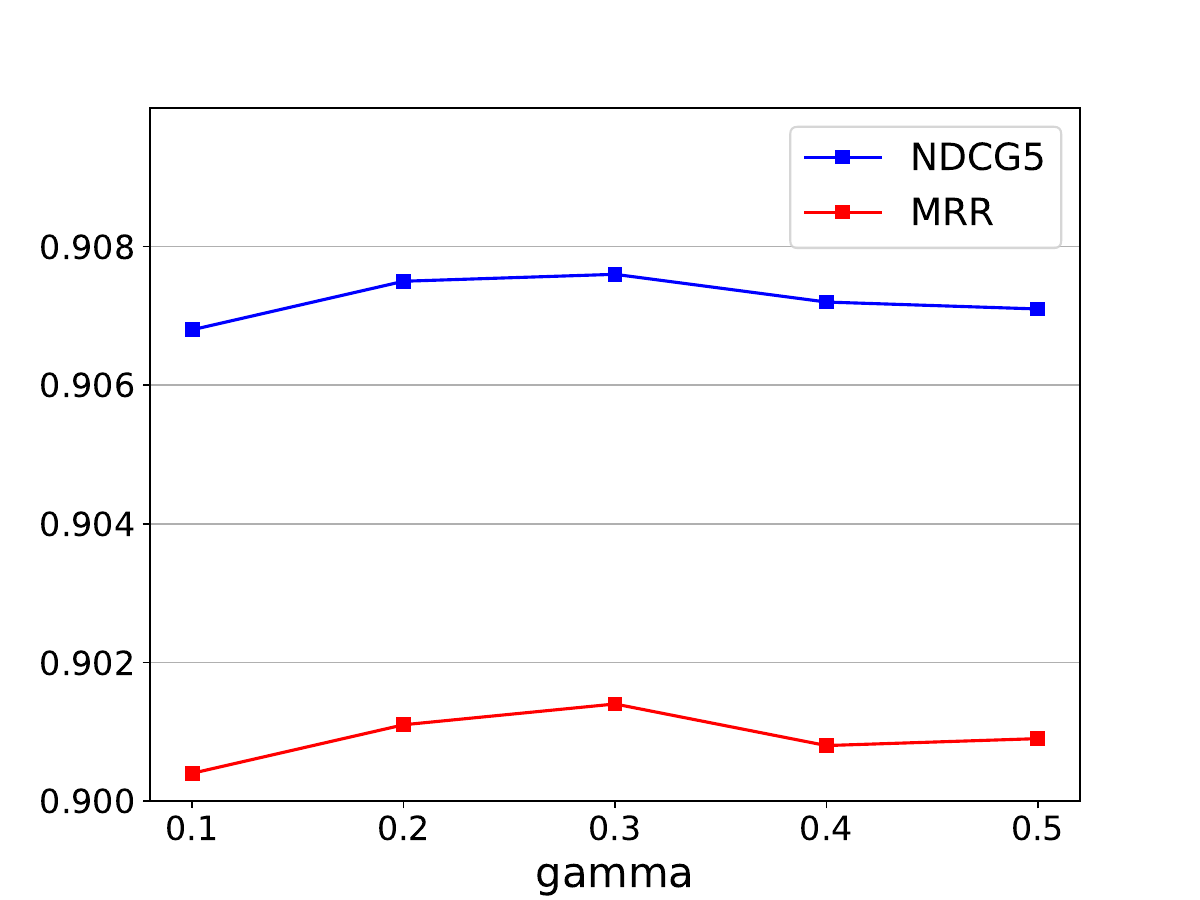}
            \caption{Junyi}
    \end{subfigure}
    \centering
    \caption{Comparing the performance across different edge masking ratio $\gamma$.}
    \label{fig:gamma}
\end{figure*}

A crucial component of our framework involves performing contrastive learning on the knowledge graph through edge dropout. A key parameter in this approach is the masking ratio of edges $\gamma$ used to generate different views. We investigate the impact of this parameter, with the experimental results displayed in Figure ~\ref{fig:gamma}.  It can be observed that the impact of this parameter is relatively minor on ASSIST12, while it has a more significant effect on both Junyi and ASSIST09 datasets. A certain drop ratio must be maintained to allow the GNN to learn the structure of the graph effectively. Therefore, when the masking ratio is low, the performance tends to be poorer.

\subsection{Embedding Space Transformation} \label{sec:embd}

\begin{figure*}[htbp]
    \centering
    \begin{subfigure}[t]{0.325\textwidth}
           \centering
           \includegraphics[width=\columnwidth]{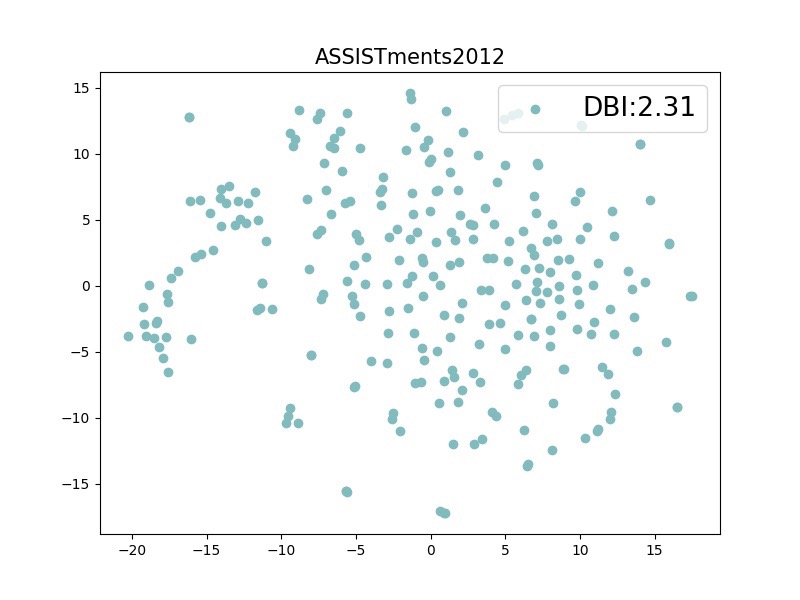}
            \caption{Vanilla Text embeddings}
    \end{subfigure}
    \begin{subfigure}[t]{0.325\textwidth}
           \centering
           \includegraphics[width=\columnwidth]{Figs/Exp_figs/embeds/ASSISTments2009_moe_embeds.pdf}
            \caption{MoE embeddings}
    \end{subfigure}
    \begin{subfigure}[t]{0.325\textwidth}
           \centering
           \includegraphics[width=\columnwidth]{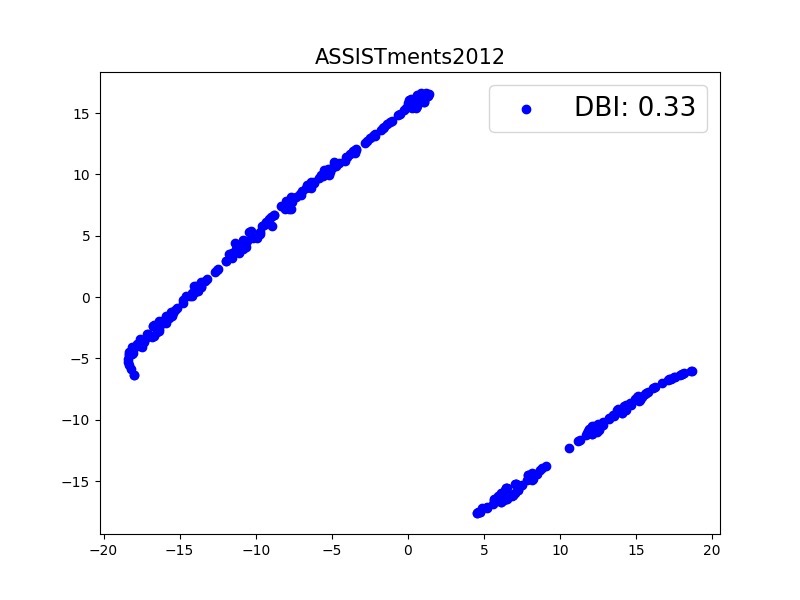}
            \caption{\our}
    \end{subfigure}
    \centering
    \caption{Comparing different distributions of embeddings of ASSIST12.The smaller the DBI value upper in figure, the better.}
    \label{fig:embedsDist12}
\end{figure*}

\begin{figure*}[htbp]
    \centering
    \begin{subfigure}[t]{0.325\textwidth}
           \centering
           \includegraphics[width=\columnwidth]{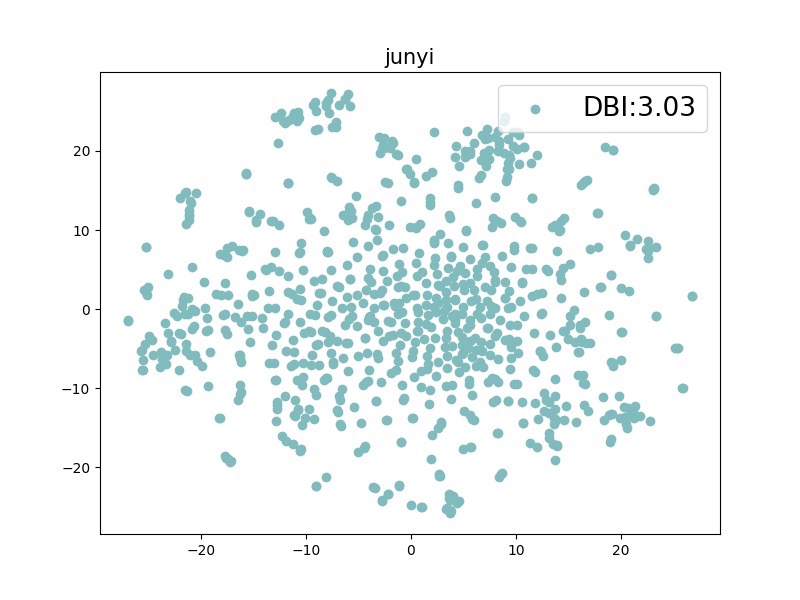}
            \caption{Vanilla Text embeddings}
    \end{subfigure}
    \begin{subfigure}[t]{0.325\textwidth}
           \centering
           \includegraphics[width=\columnwidth]{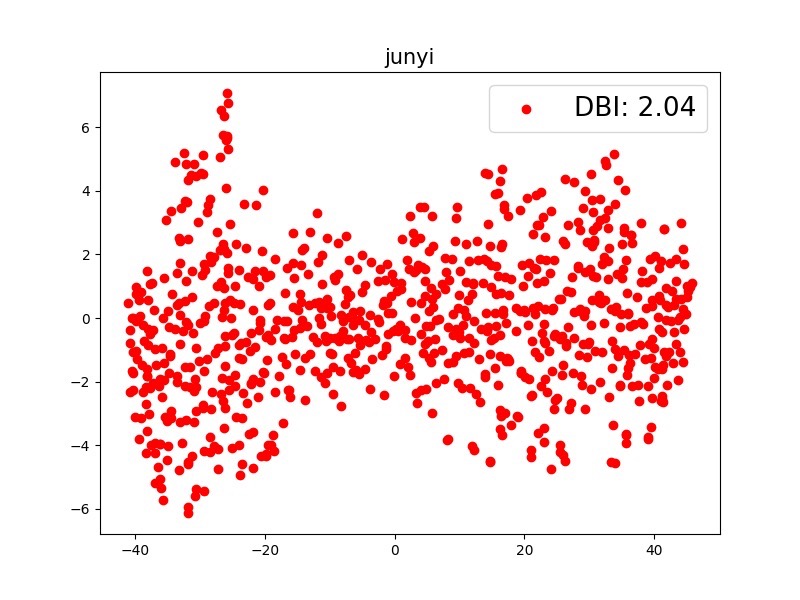}
            \caption{MoE embeddings}
    \end{subfigure}
    \begin{subfigure}[t]{0.325\textwidth}
           \centering
           \includegraphics[width=\columnwidth]{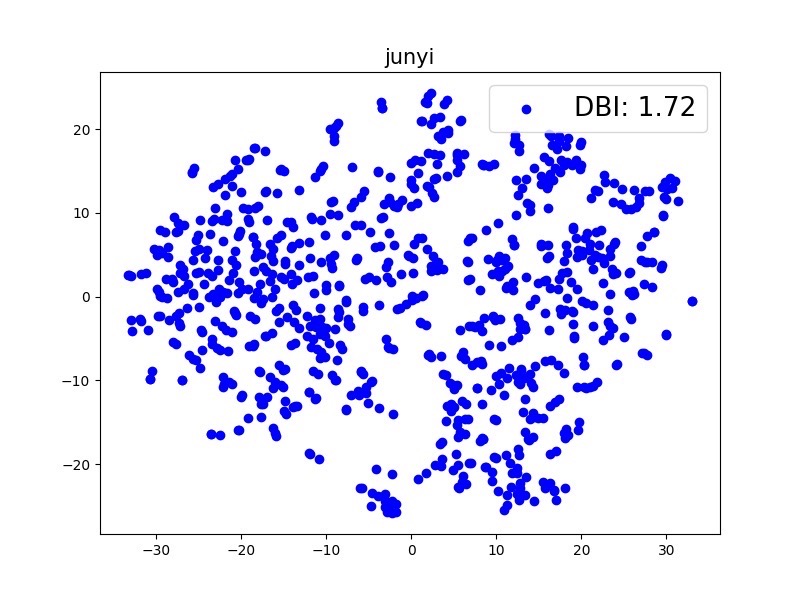}
            \caption{\our}
    \end{subfigure}
    \centering
    \caption{Comparing different distributions of embeddings of Junyi.The smaller the DBI value upper in figure, the better.}
    \label{fig:embedsDistjunyi}
\end{figure*}

In Figures ~\ref{fig:embedsDist12} and Figure ~\ref{fig:embedsDistjunyi}, we provide a comparison of the embedding distributions through different adaptations on the ASSISTments2012 and Junyi datasets, respectively. Additionally, we present their Davies-Bouldin Index (DBI) values, demonstrating that the distribution post-transformation by the graph adapter exhibits superior clustering properties compared to the MoE (Mixture of Experts) adapter.


\end{document}